\documentclass[12pt,english,onecolumn,draftnocls]{IEEEtran}
\usepackage[T1]{fontenc}
\usepackage{geometry}
\usepackage{babel}
\usepackage{array}
\usepackage{multirow}
\usepackage{amsthm}
\usepackage{amsmath}
\usepackage{amssymb}
\usepackage{graphicx}
\usepackage{setspace}
\PassOptionsToPackage{normalem}{ulem}
\usepackage{ulem}
\usepackage[unicode=true,
 bookmarks=true,bookmarksnumbered=true,bookmarksopen=true,bookmarksopenlevel=1,
 breaklinks=false,pdfborder={0 0 0},backref=false,colorlinks=false]
 {hyperref}
\hypersetup{pdftitle={Hybrid CSIT for 2-user Interference Channel},
 pdfauthor={Kaniska Mohanty},
 pdfpagelayout=OneColumn, pdfnewwindow=true, pdfstartview=XYZ, plainpages=false}

\makeatletter

\providecommand{\tabularnewline}{\\}

  \theoremstyle{definition}
  \newtheorem{defn}{\protect\definitionname}
  \theoremstyle{plain}
  \newtheorem{lem}{\protect\lemmaname}
  \theoremstyle{plain}
  \newtheorem{cor}{\protect\corollaryname}
  \theoremstyle{plain}
  \newtheorem{thm}{\protect\theoremname}
  \theoremstyle{remark}
  \newtheorem{rem}{\protect\remarkname}

\ifCLASSOPTIONcompsoc
\usepackage[caption=false,font=normalsize,labelfont=sf,textfont=sf]{subfig}
\else
\usepackage[caption=false,font=footnotesize]{subfig}
\fi

\makeatother

\providecommand{\corollaryname}{Corollary}
\providecommand{\definitionname}{Definition}
\providecommand{\lemmaname}{Lemma}
\providecommand{\remarkname}{Remark}
\providecommand{\theoremname}{Theorem}

\begin{document}

\title{The Degrees of Freedom Region of the MIMO Interference Channel with
Hybrid CSIT}

\author{Kaniska Mohanty, Chinmay S. Vaze, and Mahesh K. Varanasi~%
\thanks{The authors are with the Department of Electrical, Computer, and Energy
Engineering, University of Colorado, Boulder, CO 80309-0425, e-mail:
\protect\href{http://Kaniska.mohanty, Chinmay.Vaze, varanasi@colorado.edu}{Kaniska.mohanty, Chinmay.Vaze, varanasi@colorado.edu}.%
}}
\maketitle
\begin{abstract}
The degrees of freedom (DoF) region of the two-user MIMO (multiple-input
multiple-output) interference channel is established under a new model
termed as hybrid CSIT. In this model, one transmitter has delayed
channel state information (CSI) and the other transmitter has instantaneous
CSIT, of incoming channel matrices at the respective unpaired receivers,
and neither transmitter has any knowledge of the incoming channel
matrices of its respective paired receiver. The DoF region for hybrid
CSIT, and consequently that of $2\times2\times3^{5}$ CSIT models,
is completely characterized, and a new achievable scheme based on
a combination of transmit beamforming and retrospective interference
alignment is developed. Conditions are obtained on the numbers of
antennas at each of the four terminals such that the DoF region under
hybrid CSIT is equal to that under (a) global and instantaneous CSIT
and (b) global and delayed CSIT, with the remaining cases resulting
in a DoF region with hybrid CSIT that lies somewhere in between the
DoF regions under the instantaneous and delayed CSIT settings. Further
synergistic benefits accruing from switching between the two hybrid
CSIT models are also explored.\end{abstract}

\section{Introduction}

\IEEEPARstart{ T}{he} interference channel (IC) consists of two
transmitters that communicate with their respective receivers in spite
of the interference that each transmitter causes to its unpaired receiver.
Because the exact capacity of the IC is still not known, investigation
into the degrees of freedom (DoF), which is the pre-log factor of
the capacity, has received considerable attention in recent years.
For instance, under the assumption that the receivers have perfect
knowledge of the channel matrices, the complete DoF region of the
($M_{1},M_{2},N_{1},N_{2}$) MIMO Gaussian IC was characterized for
the instantaneous CSIT setting in \cite{DBLP:journals/tit/JafarF07},
for the no CSIT setting under the assumption on fading distributions
that transmit directions at different receiver antennas are statistically
indistinguishable in \cite{DBLP:journals/tit/HuangJSV12}, \cite{DBLP:journals/corr/abs-0909-5424,DBLP:journals/corr/abs-1008-5196,DBLP:journals/corr/abs-1105-6033},
and for the delayed CSIT setting in \cite{DBLP:journals/corr/abs-1101-5809}. 

This paper sheds light on the unexplored region between the instantaneous
CSIT and delayed CSIT settings for the MIMO IC. It is not difficult
to envisage a situation where the two transmitters have different
types of CSI. For example, via feedback links, one transmitter has
perfect and instantaneous CSIT corresponding to the incoming channels
at a stationary receiver, while the other transmitter has only delayed
CSIT corresponding to the incoming channels at a mobile receiver.
We model such a combination of perfect and instantaneous CSIT at one
transmitter and delayed CSIT at the other, under the name \textbf{Hybrid
CSIT. }

We address some of the pertinent questions for such a model, especially
the usefulness of providing perfect and instantaneous CSIT at one
transmitter. To analyze the improvement in the DoF region over that
of delayed CSIT accruing from this extra information, we first characterize
the complete DoF region for the $\left(M_{1},M_{2},N_{1},N_{2}\right)$
MIMO Gaussian IC under the hybrid CSIT model. Without loss of generality,
we assume $N_{1}\geq N_{2}$. When transmitter 1 has perfect and instantaneous
CSIT of incoming channels at receiver 2, we call the model hybrid
CSIT 1; and analogously hybrid CSIT 2 refers to the model where transmitter
2 has perfect and instantaneous CSIT of incoming channels at receiver
1. No channel state information at a transmitter is assumed of incoming
channels at its paired receiver. The DoF regions for the two models
are characterized separately. 

To show the achievability of the DoF region, we develop a new two-phase
achievability scheme. The instantaneous CSIT at one receiver is exploited
for transmit beamforming in the null space of the unpaired receiver,
thus nulling the interference at that receiver. Delayed CSIT at the
other transmitter is used for retrospective interference alignment
(RIA) in the second phase of the scheme. This combination of transmit
beamforming and RIA allows us to take maximal advantage of both forms
of CSIT.

The characterization of the DoF regions for the hybrid CSIT models
allows us to compare delayed CSIT, hybrid CSIT 1 and 2 and perfect
and instantaneous CSIT models. This yields the encouraging result
that for a large number of cases, hybrid CSIT 1 achieves the DoF region
for perfect and instantaneous CSIT. Even for those cases where it
fails to achieve the instantaneous CSIT DoF region, hybrid CSIT 1
still manages to improve over delayed CSIT. The results for hybrid
CSIT 2 are more sobering; with the exception of one case, the DoF
region remains the same as that of delayed CSIT. But for the one remaining
case, there is an improvement over delayed CSIT, and the DoF region
lies between that of delayed CSIT and perfect and instantaneous CSIT.
The intuition behind the differing behavior of the two models becomes
clearer when we realize that hybrid CSIT 1 allows interference nulling
at the more constrained receiver i.e., the one with fewer antennas,
thus allowing much bigger gains from transmit beamforming.

In an attempt at classification of CSIT models for the 2-user MIMO
IC, we introduce four other CSIT models, namely the weaker hybrid
CSIT 1 and 2 and enhanced hybrid CSIT 1 and 2 models. The weaker hybrid
CSIT models are similar to the corresponding hybrid CSIT models, with
the singular exception that the transmitter with instantaneous CSIT
has no knowledge of the direct channel of its unpaired receiver in
the weaker hybrid CSIT model. In the enhanced hybrid CSIT models,
the transmitter with the delayed information in the corresponding
original hybrid CSIT model knows its own cross-channel with delay,
and all other channels are known at all transmitters instantly. We prove
that the DoF region obtained for each of the original hybrid CSIT
models also characterizes completely the DoF region of a large class
of CSIT models, ranging from the corresponding weaker hybrid CSIT
model to the corresponding enhanced hybrid CSIT model. More precisely,
allowing the CSIT for each channel at each transmitter to be in one
of three states i.e., unknown, known with a delay or known instantaneously
and perfectly, we demonstrate that the DoF regions found in this work
for the hybrid CSIT models applies to a total of $2\times2\times3^{5}$
CSIT models. 

The DoF tuple of $\left(1,\frac{1}{2}\right)$ was shown to be achievable
for the 2-user multiple-input single-output (MISO) BC with hybrid
CSIT in \cite{DBLP:journals/jstsp/MalekiJS12}, when the transmitter
had instantaneous CSIT about one receiver and delayed CSIT about the
other receiver. The characterization of the DoF region of the 2-user
MIMO BC under the hybrid CSIT model was done later in \cite{6328519}.

Although recent work in \cite{6197205,DBLP:journals/corr/abs-1203-2550,DBLP:journals/corr/abs-1205-3474}
has investigated so-called mixed CSIT models for the two-user MISO
BC, its extension to the IC in \cite{DBLP:journals/corr/abs-1204-3046}
is fundamentally different from the hybrid CSIT models analyzed here.
The mixed CSIT model for the IC in \cite{DBLP:journals/corr/abs-1204-3046}
applies to situations where each transmitter has the same mix of both
delayed CSIT and an imperfect version of instantaneous CSIT. On the
other hand, transmitters in the hybrid CSIT models are distinguishable
by the fact that they incorporate the heterogeneity of terminals as
it relates to their mobility, the capacity of their feedback links,
etc., resulting in the knowledge of channels being \textit{different}
at the two transmitters. The focus in this paper is on having delayed
CSIT at one transmitter and instantaneous CSIT at the other regarding
the cross links and the direct links of the unpaired receivers, with
no CSIT at either transmitter of the cross and direct channels at
its respective paired receiver. We note here again that for the two-user
MISO BC, \cite{DBLP:journals/corr/abs-1205-3474} establishes the
DoF of the general mixed-CSIT model wherein the quality of current
channel knowledge is different at the two transmitters which therefore
also generalizes the work of \cite{6197205,DBLP:journals/corr/abs-1203-2550}
as well as the DoF result for the hybrid CSIT model of \cite{DBLP:journals/jstsp/MalekiJS12}. 

It has also come to the attention of the authors after this work was
posted on Arxiv (\cite{Mohanty:arXiv1209.0047}) that \cite{DBLP:journals/corr/abs-1208-5071}
considers the so-called alternating CSIT model for the two-user multiple-input
single-output (MISO) broadcast channel (BC) with a two-antenna transmitter
and two single-antenna receivers, where the transmitter\textquoteright{}s
knowledge of the channels to the two receivers alternates between
the various combinations of instantaneous, delayed and no CSIT (like
the hybrid models of this work) and this variation of channel knowledge
is shown to yield additional or synergistic DoF benefits. However,
while \cite{DBLP:journals/corr/abs-1208-5071} characterizes the DoF
of the two-user MISO BC for the alternating CSIT model, this work
characterizes the DoF of certain hybrid CSIT models for the two-user
MIMO IC with an arbitrary number of antennas at each of the four nodes.
Demonstration of such a synergistic benefit accrued by switching between
the two hybrid CSIT models is also given. Extensions of achievable
schemes in \cite{DBLP:journals/corr/abs-1208-5071} from the 2-user
MISO BC to the 2-user MISO IC are shown. The alternating CSIT model
is further developed for the MIMO IC, with an example of an achievability
scheme for the MIMO IC with synergistic benefits being illustrated.

The channel model is described in the next section, followed by main
results in Section \ref{sec:OUTER-BOUNDS}. Section \ref{sec:OUTER-BOUNDS}
also includes the proof of the outer bounds for the DoF region, as
well as an extension of these outer bounds to a much larger class
of CSIT models. A detailed example of the achievability scheme is
also given in the same section, with a discussion about the range
of CSIT models the achievability scheme applies to. Examples illustrating
the extension of this achievability scheme into the realms of alternating
CSIT are developed next. The generalized achievability scheme is given
in Section \ref{sec:ACHIEVABILITY-SCHEME}, followed by detailed proofs
of the optimality of the scheme in Section \ref{sec:Proofs-of-Main-Results}.
Conclusions can be found in Section \ref{sec:CONCLUSION}.

\section{THE CHANNEL MODEL \label{sec:THE-CHANNEL-MODEL}}

In this section, we describe the ($M_{1},M_{2},N_{1},N_{2})$ MIMO
interference channel (IC) under the assumption of hybrid CSIT. The
channel consists of two transmitters $T_{1}$ and $T_{2}$, with $M_{1}$
and $M_{2}$ antennas, respectively, and their paired receivers $R_{1}$
and $R_{2}$, with $N_{1}$ and $N_{2}$ antennas, respectively. Without
loss of generality, we assume that $N_{1}\geq N_{2}$. Here, transmitter
$T_{i}$ ($i=1,2$) has a message intended only for the receiver $R_{i}$,
but its transmit signal causes interference at the other receiver.
At the $t^{th}$channel use, signals received at the two receivers
are given by 
\begin{eqnarray*}
Y_{1}(t) & = & H_{11}(t)X_{1}(t)+H_{12}(t)X_{2}(t)+W_{1}(t)\\
Y_{2}(t) & = & H_{21}(t)X_{2}(t)+H_{22}(t)X_{2}(t)+W_{2}(t)
\end{eqnarray*}
where $X_{i}(t)\in\mathbb{C}^{M_{i}\times1}$ is the transmitted signal
from transmitter $T_{i}$; $Y_{i}(t)\in\mathbb{C}^{N_{i}\times1}$
is the signal received by receiver $R_{i}$; $W_{i}(t)$ is the additive
noise at receiver $R_{i}$; $H_{ij}(t)\in\mathbb{C}^{N_{i}\times M_{j}}$is
the channel matrix from Transmitter $T_{j}$ to Receiver $R_{i}$;
and both transmitters have a power constraint of $P$ i.e $E(||X_{i}(t)||^{2})\leq P.$

The channels are assumed to be Rayleigh faded, i.e., all entries of
all matrices ${H_{ij}(t)}$ are independent and identically distributed
(i.i.d.) zero-mean, unit-variance complex normal $\mathcal{CN}(0,1)$
random variables. The additive Gaussian noise is similarly modeled
as a complex normal random variable i.e., $W_{i}(t)\sim\mathcal{CN}(0,I_{N_{i}})$.
Also, the channel matrices and the Gaussian noise are assumed to be
i.i.d across time and independent of each other. 

Both the receivers are assumed to have perfect and, without loss of
generality, instantaneous knowledge of all channel matrices. In this
paper, we investigate the DoF region when one transmitter has perfect
and instantaneous CSIT, while the other transmitter has access only
to delayed CSIT. More precisely, we study the situation where the
channel matrices corresponding to one receiver are known at its unpaired
transmitter perfectly and instantaneously, while the channel matrices
corresponding to the other receiver are known to its unpaired transmitter
with some finite delay, assumed without loss of generality and for
convenience to be of 1 time slot. Such a situation, for instance,
can be a consequence of differing mobility of the two receivers. 

In \textbf{hybrid CSIT 1,} transmitter $T_{1}$ has perfect and instantaneous
knowledge of channel matrices $H_{21},H_{22}$ corresponding to receiver
$R_{2}$, while transmitter $T_{2}$ has delayed knowledge of the
matrices $H_{12},H_{11}$ corresponding to receiver $R_{1}$. Moreover,
transmitter $T_{1}$ has no knowledge of channels of receiver $R_{1}$
and transmitter $T_{2}$ has no knowledge of channels of receiver
$R_{2}$. In \textbf{hybrid CSIT 2}, transmitter $T_{2}$ has perfect
and instantaneous knowledge of channel matrices $H_{11},H_{12}$ corresponding
to receiver $R_{1}$ while transmitter $T_{1}$ has delayed knowledge
of $H_{21},H_{22}$ corresponding to receiver $R_{2}$. As in the
case of the Hybrid CSIT 1 model, neither transmitter has knowledge
of channels of its paired receiver. $D^{h1}$ and $D^{h2}$ refer
to the DoF region for the hybrid CSIT 1 and hybrid CSIT 2 cases, respectively.
Because of the assumption $N_{1}\geq N_{2}$, the two cases are not
symmetric and will need to be dealt with separately.
\begin{figure}[tb]
\begin{minipage}[t]{0.45\textwidth}%
\begin{center}
\includegraphics[clip,scale=0.45]{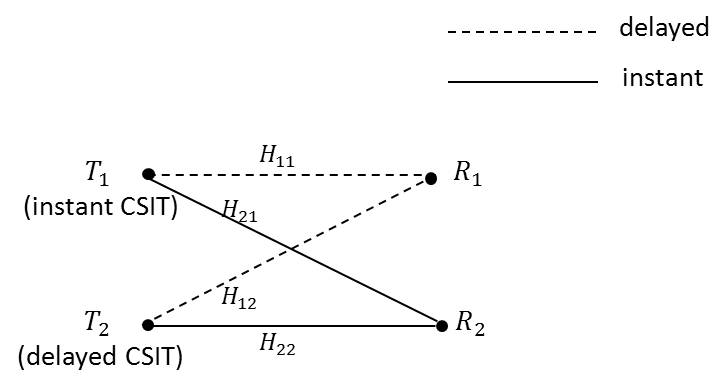}\caption{Hybrid CSIT 1 model. $T_{1}$ learns $H_{21},H_{22}$ instantly, $T_{2}$
learns $H_{12},H_{11}$ with unit delay.}

\par\end{center}%
\end{minipage}\hfill{}%
\begin{minipage}[t]{0.45\textwidth}%
\begin{center}
\includegraphics[clip,scale=0.45]{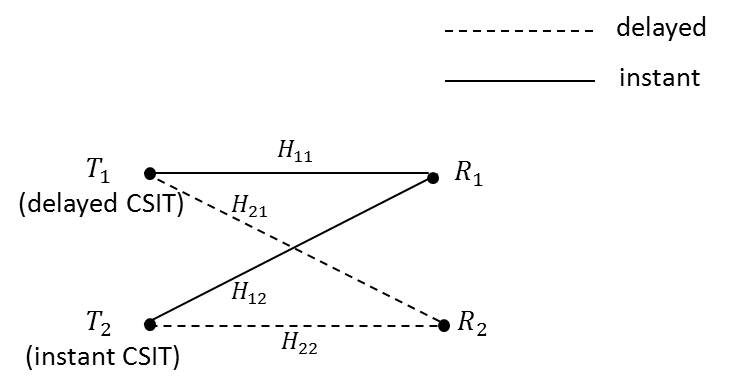}\caption{Hybrid CSIT 2 model. $T_{1}$ learns $H_{21},H_{22}$ with unit delay,
$T_{2}$ learns $H_{12},H_{11}$ instantly.}

\par\end{center}%
\end{minipage}
\end{figure}
 
\begin{figure}[tb]
\begin{minipage}[t]{0.45\textwidth}%
\begin{center}
\includegraphics[clip,scale=0.45]{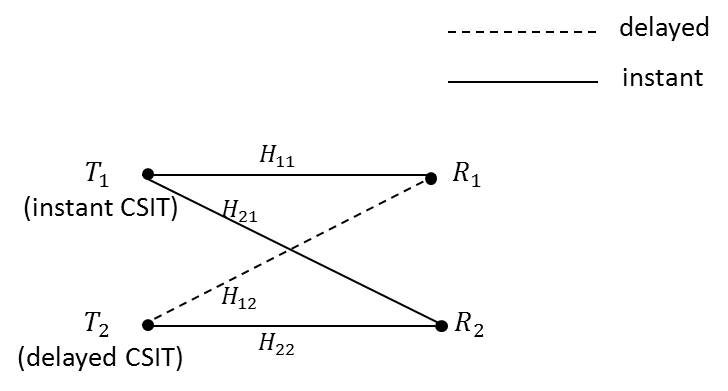}\caption{\label{fig:Enhanced-Hybrid-CSIT-1}Enhanced Hybrid CSIT 1 model. $T_{2}$
learns $H_{12}$ with unit delay, $T_{1}$ and $T_{2}$ learn all
other channels instantly.}

\par\end{center}%
\end{minipage}\hfill{}%
\begin{minipage}[t]{0.45\textwidth}%
\begin{center}
\includegraphics[clip,scale=0.45]{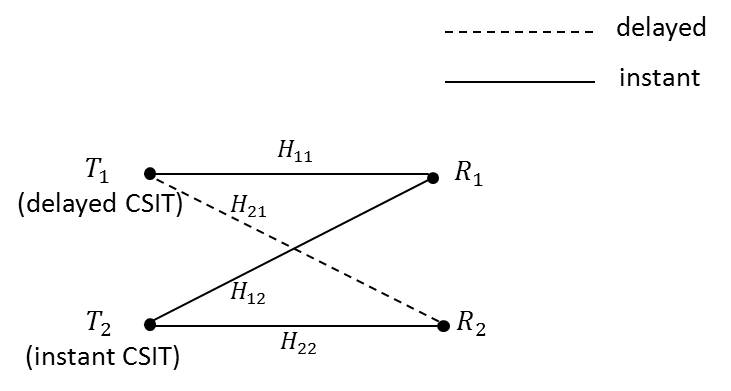}\caption{\label{fig:Enhanced-Hybrid-CSIT-2}Enhanced Hybrid CSIT 2 model. $T_{1}$
learns $H_{21}$ with unit delay, $T_{1}$ and $T_{2}$ learn all
other channels instantly.}

\par\end{center}%
\end{minipage}
\end{figure}
 
\begin{figure}[t]
\begin{minipage}[t]{0.45\textwidth}%
\begin{center}
\includegraphics[clip,scale=0.45]{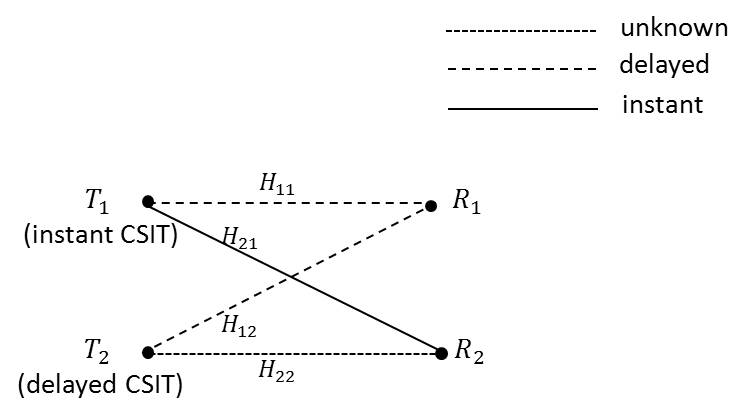}\caption{\label{fig:Weaker-Hybrid-CSIT-1}Weaker Hybrid CSIT 1 model. $T_{1}$
learns $H_{21}$ instantly, $T_{2}$ learns $H_{12},H_{11}$ with
unit delay.}

\par\end{center}%
\end{minipage}\hfill{}%
\begin{minipage}[t]{0.45\textwidth}%
\begin{center}
\includegraphics[clip,scale=0.45]{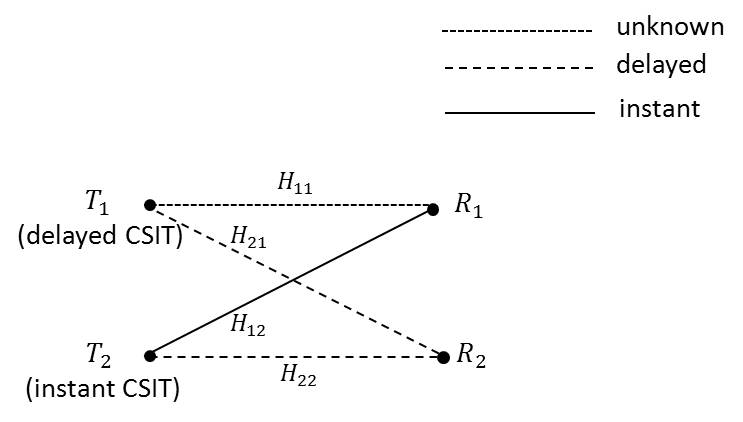}\caption{\label{fig:Weaker-Hybrid-CSIT-2}Weaker Hybrid CSIT 2 model. $T_{1}$
learns $H_{21},H_{22}$ with unit delay, $T_{2}$ learns $H_{12}$
instantly.}

\par\end{center}%
\end{minipage}
\end{figure}

Knowledge of each of the four channel matrices $H_{11},H_{12},H_{21}$
and $H_{22}$ at each of the two transmitters can be in one of three
states: unknown, known with a delay or known perfectly and instantaneously.
Thus, there exist a total of $3^{8}$ possibilities for side information.
Fortunately, not all these possibilities require individual attention.
Indeed, it is shown in this work that the DoF region of the hybrid
CSIT 1 model applies to $2\times3^{5}$ side information models as
does the DoF region of the hybrid CSIT 2 model. We note here that
the DoF region obtained in \cite{DBLP:journals/corr/abs-1101-5809}
for the delayed CSIT model applies to $2\times2\times3^{4}$ possible
side information models and the DoF region of the perfect and instantaneous
CSIT model in \cite{DBLP:journals/tit/JafarF07} similarly applies
to another $3^{6}$ side information models. Although a complete classification
of all the $3^{8}$ models is not available at this time, this paper
makes progress in that direction.

To illustrate the range of CSIT models to which the DoF regions obtained
in this paper are applicable, we define four other CSIT models : \textbf{enhanced
hybrid CSIT} 1 and 2 models, shown in Fig. \ref{fig:Enhanced-Hybrid-CSIT-1}
and \ref{fig:Enhanced-Hybrid-CSIT-2}, and \textbf{weaker hybrid CSIT}
1 and 2 models, shown in Fig. \ref{fig:Weaker-Hybrid-CSIT-1} and
\ref{fig:Weaker-Hybrid-CSIT-2}. In the weaker hybrid CSIT models,
the transmitter with perfect and instantaneous CSIT does not know
the direct channel of its unpaired receiver, and the knowledge about
the rest of the channels at each receiver remains the same as in the
respective original hybrid CSIT model. In the weaker hybrid CSIT 1
model, $T_{1}$ knows instantaneously the cross-channel matrix $H_{21}$
whereas $T_{2}$ has delayed knowledge of $H_{12}$ and $H_{11}$.
Similarly, in the weaker hybrid CSIT 2 model, $T_{2}$ knows instantaneously
the cross-channel matrix $H_{12}$ whereas $T_{1}$ has delayed knowledge
of $H_{21}$ and $H_{22}$. In particular, $T_{1}$ has no knowledge
of $H_{22}$ in the weaker hybrid CSIT 1 model and $T_{2}$ has no
knowledge of $H_{11}$ in the weaker hybrid CSIT 2 model. In the enhanced
hybrid CSIT 1 model, the cross-channel $H_{12}$ is known with a delay
at $T_{2}$ and instantaneously at $T_{1}$, and all other channels
are known at both transmitters perfectly and instantaneously. Similarly,
in the enhanced version of the hybrid CSIT 2 model, $H_{21}$ is known
with a delay at $T_{1}$ and known instantaneously at $T_{2}$, and
all other channels are known at both receivers perfectly and instantaneously.

Let $\mathcal{M}_{1}$ and $\mathcal{M}_{2}$ be the two independent
messages to be sent from $T_{1}$ to $R_{1}$ and $T_{2}$ to $R_{2}$
respectively. A rate tuple $\left((R_{1}(P),R_{2}(P)\right)$ is said
to be achievable if there exists a codeword spanning $n$ channel
uses, with a power constraint of $P$, such that the probability of
error at both receivers goes to zero as $n\rightarrow\infty$, where
$R_{i}(P)=\log(|\mathcal{M}_{i}|)/n$. The capacity region $C(P)$
of the IC is the region of all such achievable rate tuples, and the
degree of freedom is defined as the pre-log factor of the capacity
region i.e.,
\[
\mathbf{D}=\biggl\{(d_{1},d_{2})\biggl|\ d_{i}\geq0\ {\rm and}\ \exists\ \left(R_{1}(P),R_{2}(P)\right)\in C(P)
\]
\begin{equation}
\left.\text{\text{such that}}\ d_{i}=\lim_{P\rightarrow\infty}\frac{R_{i}(P)}{\log(P)}\:,i\in\{1,2\}\right\} .\label{eq:DoF-region-Defintion}
\end{equation}

The DoF region of various scenarios that are referred to in this paper
are as follows:
\begin{description}
\item [{$D^{no}$}] Neither transmitter has any knowledge of channel realizations.
\item [{$D^{d}$}] Both transmitters have delayed CSIT i.e., $T_{1}$ knows
$H_{21},H_{22}$ and $T_{2}$ knows $H_{12},H_{11}$, both with a
delay. This DoF region has been characterized in \cite{DBLP:journals/corr/abs-1101-5809}.
\item [{$D^{h1}$}] $T_{1}$ has perfect and instantaneous knowledge of
$H_{21},H_{22}$; while $T_{2}$ has delayed CSI of $H_{12},H_{11}$.
\item [{$D^{h2}$}] $T_{2}$ has perfect and instantaneous knowledge of
$H_{12},H_{11}$; $T_{1}$ has delayed CSI of $H_{21},H_{22}$.
\item [{$D^{i}$}] Both transmitters have perfect and instantaneous knowledge
of the channel matrices of their unpaired receivers. This DoF region
has been characterized in \cite{DBLP:journals/tit/JafarF07}.
\end{description}

\section{Main Results\label{sec:OUTER-BOUNDS}}
\begin{defn}
For $i\in\left\{ 1,2\right\} $, Condition $i$ is said to hold, whenever
the inequality 
\[
M_{i}>N_{1}+N_{2}-M_{j}>N_{i}>N_{j}>M_{j}>N_{j}\frac{N_{j}-M_{j}}{N_{i}-M_{j}}
\]
is true for $j\in\left\{ 1,2\right\} $ with $j\ne i$. The two conditions
can not be true simultaneously, and are symmetric counterparts of
each other. Also, condition $i$ can not hold if $N_{j}\geq N_{i}$.
\end{defn}
We list all the outer bounds from \cite{DBLP:journals/corr/abs-1101-5809}
that are applicable to the more constrained delayed CSIT case in inequalities
\ref{eq:1}-\ref{eq:7}.
\begin{eqnarray}
L_{o1} & \equiv & 0\leq d_{1}\leq\min(M_{1},N_{1});\label{eq:1}\\
L_{o2} & \equiv & 0\leq d_{2}\leq\min(M_{2},N_{2});\\
L_{1} & \equiv & \frac{d_{1}}{\min(N_{1}+N_{2},M_{1})}+\frac{d_{2}}{\min(N_{2},M_{1})}\leq\frac{\min(N_{2},M_{1}+M_{2})}{\min(N_{2},M_{1})};\\
L_{2} & \equiv & \frac{d_{1}}{\min(N_{1},M_{2})}+\frac{d_{2}}{\min(N_{1}+N_{2},M_{2})}\leq\frac{\min(N_{1},M_{1}+M_{2})}{\min(N_{1},M_{2})};\\
L_{3} & \equiv & d_{1}+d_{2}\leq\min\left[M_{1}+M_{2},N_{1}+N_{2},\max(M_{1},N_{2}),\max(M_{2},N_{1})\right];\\
L_{4} & \equiv & d_{1}+d_{2}\frac{N_{1}+2N_{2}-M_{2}}{N_{2}}\leq N_{1}+N_{2},{\rm if\ condition\ 1\ holds};\\
L_{5} & \equiv & d_{2}+d_{1}\frac{N_{2}+2N_{1}-M_{1}}{N_{1}}\leq N_{1}+N_{2},{\rm if\ condition\ 2\ holds}.\label{eq:7}
\end{eqnarray}

\begin{lem}
The outer bounds $L_{2}$ and $L_{5}$ apply to the hybrid CSIT 1
model, while the outer bounds $L_{1}$ and $L_{4}$ apply to hybrid
CSIT 2. The outer bounds $L_{01},L_{02},L_{3}$ hold for both hybrid
CSIT 1 and CSIT 2 models.\label{Lemma 3}\label{Lemma 2}\label{Lemma 1}\end{lem}
\begin{IEEEproof}
The outer bounds $L_{1},L_{2},L_{4}$ and $L_{5}$ have been proved
for the delayed CSIT model in \cite{DBLP:journals/corr/abs-1101-5809}.
We first give a quick summary of the major elements of the proof of
$L_{1}$ for the delayed CSIT model, before pointing out why $L_{1}$and
$L_{4}$ do not apply to the hybrid CSIT 1 model. In \cite{DBLP:journals/corr/abs-1101-5809},
under the assumption that $R_{1}$ does not face any interference,
it was first shown that it is sufficient to consider only the first
$M_{1}-M_{2}$ and $M_{2}$ antennas of $R_{1}$ and $R_{2}$ respectively
(Lemmas 2 and 3 in \cite{DBLP:journals/corr/abs-1101-5809}). The
statistical equivalence of channel outputs was then established, which
states that the signals received at any two antennas at time $t$
are statistically equivalent to each other, conditioned on the received
outputs at previous times, knowledge of all present and past channels
and the outputs at some other antennas at time $t$ (Lemma 4 and Remark
7 in \cite{DBLP:journals/corr/abs-1101-5809}). These results were
used to derive various inequalities involving the differential entropy
of the signals at the two receivers, which were in turn employed to
establish a lower bound on the DoF occupied by the interference at
$R_{2}$ (Lemma 1 in \cite{DBLP:journals/corr/abs-1101-5809}), which
was shown to be proportional to $d_{1}$. Finally, the outer bound
$L_{1}$ for the delayed CSIT case was obtained from Fano's inequality
and the above mentioned lower bound on the interference DoF at $R_{2}$.
The outer bound $L_{4}$ for the delayed CSIT case was also proved
using the same techniques involved in the proof $L1$, notably the
statistical equivalence of channel outputs, but while accounting for
the interference seen at both receivers.

A more detailed perusal of the proofs of $L_{1}$ and $L_{4}$, in
particular Lemma 4 in \cite{DBLP:journals/corr/abs-1101-5809}, shows
that the statistical equivalence of two antennas is based on the transmit
signal $X_{1}\left(t\right)$ being independent of $H_{2i1}\left(t\right)$
and $H_{2j1}\left(t\right)$ (where $i$ and $j$ are arbitrary antennas
of $R_{2}$, while $H_{2i1}\left(t\right)$ and $H_{2j1}\left(t\right)$
are the $i^{th}$ and $j^{th}$ rows of matrix $H_{21}\left(t\right)$),
conditioned on a set of random variables that involve past channel
outputs, present channel outputs at some other antennas and the previous
channel matrices. This condition on $X_{1}\left(t\right)$ is equivalent
to the condition that $T_{1}$ does not know the channel $H_{21}$
instantaneously, ensuring that $X_{1}$ remains uncorrelated with
$H_{2i1}\left(t\right)$ and $H_{2j1}\left(t\right)$ conditioned
on the previous set of random variables. Thus, $L_{1}$ and $L_{4}$
do not apply to the hybrid CSIT 1 model, where $T_{1}$ has instantaneous
knowledge of $H_{21}$. On the other hand, the proofs for $L_{2}$
and $L_{5}$ in \cite{DBLP:journals/corr/abs-1101-5809} , which are
the same the proofs of $L_{1}$ and $L_{4}$ respectively, except
for an exchange of roles between $R_{1}$ and $R_{2}$, are verified
to carry over for the hybrid CSIT 1 model. The analogous condition
for the statistical equivalence of channel outputs to hold in this
case is for $H_{12}$ not be known at $T_{2}$ instantaneously , a
condition that is true for the hybrid CSIT 1 model. Thus $L_{2}$
and $L_{5}$ apply to the hybrid CSIT 1 model. Similar arguments show
that $L_{1}$ and $L_{4}$ apply to the hybrid CSIT 2 model, while
$L_{2}$ and $L_{5}$ do not.

Outer bound $L_{3}$ is proved in \cite{DBLP:journals/tit/JafarF07},
where both transmitters have perfect and instantaneous CSIT. It therefore
applies to both the hybrid CSIT models under consideration. $L_{01}$
and $L_{02}$ follow from the fact that the DoF of a point-to-point
MIMO link is limited by the minimum number of transmit and receive
antennas. Since bounds $L_{01}$ and $L_{02}$ hold for every case
under consideration, they will not be explicitly mentioned in any
of the proofs for convenience. \end{IEEEproof}
\begin{cor}
\label{cor:Strengthen-outer-bounds}The outer bounds $L_{2}$ and
$L_{5}$ also apply to the enhanced hybrid CSIT 1 model and the outer
bounds $L_{1}$ and $L_{4}$ also apply to the enhanced hybrid CSIT
2 model.\end{cor}
\begin{IEEEproof}
The same argument as in Lemma \ref{Lemma 2} shows that the only condition
needed for $L_{2}$ and $L_{5}$ to hold is for $T_{2}$ to have a
delayed knowledge of $H_{12}$, which is precisely the condition mentioned
for the enhanced version of the hybrid CSIT 1 model. A similar argument
shows that $L_{1}$ and $L_{4}$ apply to the corresponding enhanced
version of the hybrid CSIT 2 model. Thus, the outer bounds derived
here are applicable to a much larger class of CSIT models than the
hybrid CSIT models.\end{IEEEproof}
\begin{defn}
$D_{outer}^{h1}$ is defined as the $\left(d_{1},d_{2}\right)$ region
bounded by $L_{01},L_{02},L_{2},L_{3}$ and $L_{5}$. Similarly, $D_{outer}^{h2}$
is defined as the $\left(d_{1},d_{2}\right)$ region bounded by $L_{01},L_{02},L_{1},L_{3}$
and $L_{4}$.\end{defn}
\begin{thm}[Hybrid CSIT 1]
The DoF region of the MIMO IC under the hybrid CSIT 1 model is equal
to $D_{outer}^{h1}$ i.e., $D^{h1}=D_{outer}^{h1}$. \label{Theorem 5}\end{thm}
\begin{IEEEproof}
Lemma \ref{Lemma 2} proves that the DoF region is outer bounded by
$D_{outer}^{h1}$ i.e., $D^{h1}\subseteq D_{outer}^{h1}$. We now
show that the $D_{outer}^{h1}$ region is achievable i.e., $D_{outer}^{h1}\subseteq D^{h1}$,
and hence $D^{h1}=D_{outer}^{h1}$.

Motivated by the proof of Theorem 2 in \cite{DBLP:journals/corr/abs-1101-5809},
we start by enumerating all possible cases of the $\left(M_{1},M_{2},N_{1},N_{2}\right)$
tuple in Table \ref{tab:Active-OUTER-BOUNDS}. The cases are exhaustive
and mutually exclusive. Except for the case $M_{1},M_{2}>N_{1},N_{2}$
(\textbf{A.I.3b} in Table \ref{tab:Active-OUTER-BOUNDS}), we show
that the DoF region for perfect and instantaneous CSIT is achievable
even with hybrid CSIT 1, using techniques from \cite{DBLP:journals/tit/JafarF07}
and \cite{DBLP:journals/corr/abs-1101-5809}. This proves that $D^{h1}=D_{outer}^{h1}=D^{i}$,
except when $M_{1},M_{2}>N_{1},N_{2}$. A detailed case-by-case analysis
of achievability is deferred to Section \ref{sec:Proofs-of-Main-Results}.
\begin{table*}[tb]
\caption{\label{tab:Active-OUTER-BOUNDS} CASE BY CASE COMPARISON OF DOF REGIONS
OF HYBRID CSIT 1 AND 2, DELAYED CSIT AND INSTANTANEOUS CSIT}

\begin{centering}
 \resizebox{\textwidth}{!}{
\begin{tabular}{|c|c|c|c|c|}
\hline 
\multicolumn{1}{|c||}{\multirow{1}{*}{Case}} & \multicolumn{1}{c||}{Definition ($N_{1}\geq N_{2}$)} & Active Bounds for Delayed CSIT (from \cite{DBLP:journals/corr/abs-1101-5809}) & Hybrid CSIT 1 & Hybrid CSIT 2\tabularnewline
\hline 
0 & $N_{2}\geq$$M_{1}$ & $L_{\{3\}}$ & $D^{d}=D^{h1}=D^{i}$ & $D^{d}=D^{h2}=D^{i}$\tabularnewline
\hline 
\multicolumn{5}{|c}{Case A: $M_{1}>N_{2}$ and $M_{2}\geq N_{2}$}\tabularnewline
\hline 
\hline 
A.I & $M_{2}\geq N_{1}$ & $L_{\{1,2\}}$ &  & \tabularnewline
\hline 
1 & $M_{1}\leq N_{1}$ & $L_{1}$ & $D^{d}\subset D^{h1}=D^{i}$ & $D^{d}=D^{h2}\subset D^{i}$\tabularnewline
\hline 
2 & $M_{1}>N_{1}$ and $M_{2}=N_{2}$ & $L_{2}$ or $L_{3}$ & $D^{d}=D^{h1}=D^{i}$ & $D^{d}=D^{h2}=D^{i}$\tabularnewline
\hline 
3a & $M_{1}>N_{1}$, $M_{2}>N_{2}$ and $M_{2}=N_{1}$ & $L_{\{1,2\}}$ & $D^{d}\subset D^{h1}=D^{i}$ & $D^{d}=D^{h2}\subset D^{i}$\tabularnewline
\hline 
3b & $M_{1}>N_{1}$, $M_{2}>N_{2}$ and $M_{2}>N_{1}$ & $L_{\{1,2\}}$ & $D^{d}\subset D^{h1}\subset D^{i}$ & $D^{d}\subset D^{h2}\subset D^{i}$\tabularnewline
\hline 
A.II & $M_{2}<N_{1}$ & $L_{\{1,3\}}$ & $D^{d}\subset D^{h1}=D^{i}$ & $D^{d}=D^{h2}\subset D^{i}$\tabularnewline
\hline 
\multicolumn{5}{|c}{Case B: $M_{1}>N_{2}$ and $N_{2}>M_{2}$}\tabularnewline
\hline 
\hline 
B & \multicolumn{1}{c||}{All cases} & $L_{\{1,3,4\}}$ & $D^{d}\subset D^{h1}=D^{i}$ & $D^{d}=D^{h2}\subset D^{i}$\tabularnewline
\hline 
\end{tabular}}
\par\end{centering}

\centering{}\medskip{}
$L_{01}$ and $L_{02}$ implicitly hold for every case under consideration.
\end{table*}

For the remaining, and most interesting, case where $M_{1},M_{2}>N_{1},N_{2}$;
we develop a new achievability scheme based on linear beamforming
and retrospective interference alignment. This achievability scheme
allows us to achieve the DoF region $D_{outer}^{h1}$ and hence is
DoF-optimal. The details of the achievability scheme are described
in Section \ref{sec:ACHIEVABILITY-SCHEME}. The key ideas behind the
scheme are illustrated with an example in the next subsection. \end{IEEEproof}
\begin{thm}[Hybrid CSIT 2]
The DoF region of the MIMO IC under the hybrid CSIT 2 model is equal
to $D_{outer}^{h2}$ i.e., $D^{h2}=D_{outer}^{h2}$. \label{Theorem 4}\end{thm}
\begin{IEEEproof}
Lemma \ref{Lemma 2} proves that the DoF region is outer bounded by
$D_{outer}^{h2}$ i.e., $D^{h2}\subseteq D_{outer}^{h2}$. We now
show that the $D_{outer}^{h2}$ region is achievable, which implies
that $D_{outer}^{h2}\subseteq D^{h2}$, and hence $D^{h2}=D_{outer}^{h2}$.

All possible cases of the $\left(M_{1},M_{2},N_{1},N_{2}\right)$
tuple are tabulated in Table \ref{tab:Active-OUTER-BOUNDS}. Except
for the case $M_{1},M_{2}>N_{1},N_{2}$ (\textbf{A.I.3b} in Table
\ref{tab:Active-OUTER-BOUNDS}), we show that the outer bound on the
DoF region for hybrid CSIT 2 is the same as the DoF region for delayed
CSIT i.e., $D_{outer}^{h2}=D^{d}$. Since $D^{d}$ is already achievable
with delayed CSIT achievability schemes from \cite{DBLP:journals/corr/abs-1101-5809},
this proves that $D_{outer}^{h2}$ is also achievable. Thus, $D^{h2}=D_{outer}^{h2}=D^{d}$,
except for when $M_{1},M_{2}>N_{1},N_{2}$. A detailed case-by-case
analysis of achievability is deferred to Section \ref{sec:Proofs-of-Main-Results}.

The achievability scheme developed previously in Theorem \ref{Theorem 5}
can be easily adapted to the hybrid CSIT 2 model for the remaining
case where $M_{1},M_{2}>N_{1},N_{2}$. This achievability scheme allows
us to achieve the DoF region $D_{outer}^{h2}$ and hence is DoF-optimal.
As mentioned previously, the details of the achievability scheme are
described in Section \ref{sec:ACHIEVABILITY-SCHEME}.\end{IEEEproof}
\begin{cor}
\label{cor:Strengthen-achievability-scheme}The achievability scheme
developed here also applies to the weaker version of the hybrid CSIT
models.\end{cor}
\begin{IEEEproof}
We recall that the transmitter with perfect and instantaneous CSIT
does not know the direct channel of its unpaired receiver in the weaker
hybrid CSIT models. A careful perusal of the example given in \ref{sub:Example-of-Achievability}
as well as the generalized alignment scheme in Section \ref{sec:ACHIEVABILITY-SCHEME}
shows that for both the hybrid CSIT models, the transmitter with the
perfect and instantaneous CSIT never needs to know the interference
it creates (if any) at its unpaired receiver. For example, transmitter
$T_{1}$ in hybrid CSIT 1 never needs to know any interference caused
at $R_{2}$. Hence, there is no necessity for that transmitter to
know the incoming channels at its unpaired receiver, since its transmission
scheme does not depend on this knowledge i.e., $T_{1}$ does not need
to know $H_{22}$ for the hybrid CSIT 1 model. Thus, the achievability
scheme applies to a weaker version of the hybrid CSIT models. \end{IEEEproof}
\begin{rem}
Corollaries \ref{cor:Strengthen-outer-bounds} and \ref{cor:Strengthen-achievability-scheme}
together establish that the DoF region characterized for each of the
two hybrid CSIT models actually holds for a total of $2\times3^{5}$
CSIT models. For example, provided $H_{21}$ is known perfectly and
instantaneously at $T_{1}$, $H_{12}$ is known with a delay at $T_{2}$
and $H_{11}$ is known either instantaneously or with a delay at $T_{2}$,
Corollaries \ref{cor:Strengthen-outer-bounds} and \ref{cor:Strengthen-achievability-scheme}
together imply that the DoF region remains the same as that of hybrid
CSIT 1. Thus, CSIT about each of the channel matrices $H_{11}$, $H_{12}$
and $H_{22}$ at $T_{1}$ and $H_{21}$ and $H_{22}$ at $T_{2}$
can be in one of three possible states i.e., not known, known with
a delay or known instantaneously, without affecting the DoF region.
These considerations, combined with the $2$ different CSIT states
for $H_{11}$ at $T_{2}$ show that the total number of CSIT models
for which the DoF region is completely characterized by the hybrid
CSIT 1 DoF region is $2\times3^{5}$. Similarly, the DoF region for
the hybrid CSIT 2 model characterizes completely the DoF regions for
another $2\times3^{5}$ CSIT models. Thus, this paper characterizes
completely the DoF region of $2\times2\times3^{5}$ constituent CSIT
models, out of a total of $3^{8}$ possible CSIT models. 
\end{rem}

\begin{rem}
Of the constituent CSIT models for which the DoF regions are known,
we note that \cite{DBLP:journals/tit/JafarF07}, which obtains the
DoF region for the perfect and global CSIT problem, actually characterizes
the DoF region for $3^{6}$ constituent CSIT models. The achievability
of the DoF region obtained therein applies to the case of side information
corresponding to each transmitter knowing instantaneously the cross-channel
into the unpaired receiver, with the knowledge about the rest of the
channel matrices at each transmitter being in any one of three possible
states i.e., not known, known with a delay or known instantaneously.
This implies of course that the DoF region found in \cite{DBLP:journals/tit/JafarF07}
for the perfect and global CSIT case where both transmitters know
all four channel matrices instantaneously actually applies to a total
of $3^{6}$ CSIT models. In the same vein, the DoF region for delayed
CSIT in \cite{DBLP:journals/corr/abs-1101-5809} applies for $2\times2\times3^{4}$
constituent CSIT models. For these constituent models, the DoF region
is characterized by the cross-channels $H_{21}$ and $H_{12}$ being
known with a delay at $T_{1}$ and $T_{2}$ respectively, $H_{22}$
and $H_{11}$ being known either with a delay or instantaneously at
$T_{1}$ and $T_{2}$ respectively, and the rest of the channel matrices
at each transmitter being in one of the three possible states, thus
giving a total of $2\times2\times3^{4}$ constituent CSIT models.
\end{rem}

\subsection{Example of Achievability Scheme\label{sub:Example-of-Achievability}}

\begin{figure*}[t]
\includegraphics[scale=0.55]{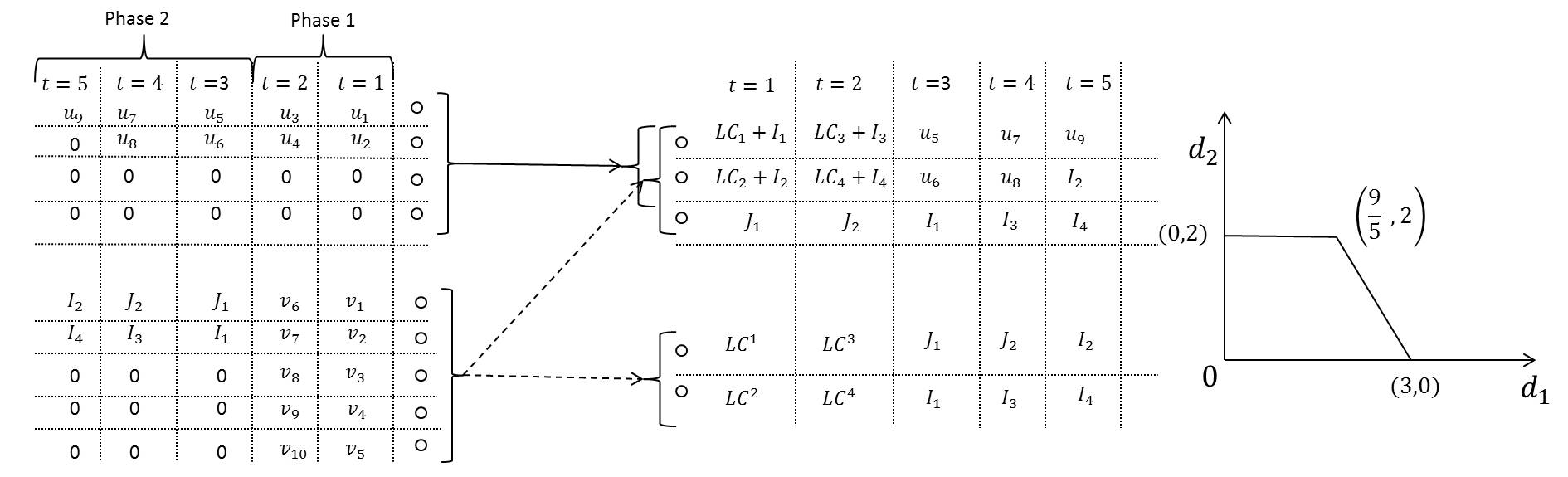}\caption{Interference alignment scheme for achieving $\left(\frac{9}{5},2\right)$
DoF pair over the $(M_{1},M_{2},N_{1},N_{2})=(4,5,3,2)$ MIMO IC under
hybrid CSIT 1 model.\label{fig:Alignment-Scheme-for}}
\end{figure*}

We present here an illustrative example of our achievable scheme for
the specific case of $\left(M_{1},M_{2},N_{1},N_{2}\right)=\left(4,5,3,2\right)$.
We consider the hybrid CSIT 1 model for this example. From Theorem
\ref{Theorem 5}, the shape of the DoF region for this choice of transmitter
and receive antennas is shown in Fig. \ref{fig:Alignment-Scheme-for}.
It is clear that if we can achieve the DoF pair $\left(\frac{9}{5},2\right)$,
then the complete DoF region can be achieved by time-sharing. To achieve
this DoF pair, it is sufficient if $T_{1}$ sends $9$ data symbols
(DS) to $R_{1}$, and $T_{2}$ sends $10$ DSs to $R_{2}$, over a
total of $5$ time extensions. The example demonstrates how this $(9,10)$
DoF pair can be achieved over $5$ time slots, under hybrid CSIT 1.
The DSs to be transmitted from $T_{1}$ are denoted as $u_{1},...,u_{9}$,
and the DSs from $T_{2}$ are $v_{1},...,v_{10}$.

\textbf{\uline{Phase 1}} 

Since additive noise does not change the DoF region, we ignore it
henceforth. We divide our scheme into two phases. In phase 1, $T_{2}$
uses all its antennas to send all the DSs it needs to send to $R_{2}$.
Since, in this example, $T_{2}$ has to send $10$ DSs and has $5$
antennas, phase 1 requires $2$ time slots to be able to send all
of them. Thus, phase 1 lasts for time slots $t=1,2$, and $v_{1},...,v_{10}$
are transmitted by $T_{2}$ in bundles of $5$ DSs per time slot over
phase 1, as seen in Fig. \ref{fig:Alignment-Scheme-for}.

In the hybrid CSIT 1 model, $T_{1}$ has instantaneous knowledge of
$H_{21}$ which allows it to do transmit beamforming in the null space
of $H_{21}$. $H_{21}$ is $2\times4$ matrix and thus has a null
space of dimension $2$. $T_{1}$ sends $2$ DSs over this 2-dimensional
null space during each time slot in phase $1$. Thus, $T_{1}$ transmits
$u_{1},u_{2}$ at time $t=1$, and $u_{3},u_{4}$ at $t=2$, in the
null space of $H_{21}$, as seen in Fig. \ref{fig:Alignment-Scheme-for}.
Evidently, these DSs cause no interference at $R_{2}$. 

Knowledge of $H_{11}(t)$ at $R_{1}$ permits it to calculate a unitary
transformation $U_{1}(t)$ matrix which allows $R_{1}$ to do receive
beam-forming. This receive beamforming is done such that the $2$
DSs sent from $T_{1}$ affect only the first $2$ antennas of $R_{1}$,
or in other words, the last row of $U_{1}\left(t\right)H_{11}\left(t\right)$
has all zeros. Over each time slot in phase 1, $R_{1}$ sees two linear
combinations of the DSs transmitted from $T_{1}$ at its first two
antennas, combined with interference from $T_{2}$. The third antenna
sees only interference from $T_{2}$. For example, in time slot $t=1$,
the first two antennas of $R_{1}$ see linear combinations $LC_{1}$
and $LC_{2}$ of $u_{1}$ and $u_{2}$, respectively, combined with
interference $I_{1}$ and $I_{2}$. The third antenna now sees interference
$J_{1}$. We note that the interference at $R_{1}$e.g., $I_{1}$,
$I_{2}$ and $J_{1}$ at $R_{1}$ are linear combinations of DSs ,
in this case $v_{1},\dots,v_{5}$, intended for $R_{2}$. 

As already stated, $R_{2}$ sees no interference because of transmit
beamforming at $T_{1}$. Thus, in phase 1, $R_{2}$ obtains two linear
combinations (of $v_{1},\dots,v_{5}$) denoted as $LC^{1}$ and $LC^{2}$
at $t=1$ and two linear combinations (of $v_{6},\dots,v_{10}$) denoted
as $LC^{3}$ and $LC^{4}$ at $t=2$. Since all the channel matrices
are Rayleigh faded and can be shown to be full-rank almost surely,
all the linear combinations and interference symbols found at both
the receivers are linearly independent with probability $1$. At the
end of phase 1, none of the receivers is able to decode its intended
data symbols. $R_{2}$ has only $4$ linear combinations available
of its $10$ intended DSs. $R_{1}$ is not able to decode the linear
combinations of its desired symbols $LC_{1},...,LC_{4}$, since they
are combined with interference $I_{1},...,I_{4}$, respectively, but
is able to learn the interference symbols $J_{1}$ and $J_{2}$.

\textbf{\uline{Phase 2}}

The objective of phase 2 is three-fold, (i) to provide sufficient
number of independent linear combinations of $v_{1},...,v_{10}$ at
$R_{2}$ so that it is able to decode all its intended DSs, (ii) to
transmit the remaining DSs $u_{5},...,u_{9}$ from $T_{1}$ and (iii)
to provide the interference symbols $I_{1},...,I_{4}$ at $R_{1}$,
which would allow it to cancel the interference and thereby access
the linear combinations $LC_{1},...,LC_{4}$. In phase 2, $T_{2}$
has (delayed) knowledge of channel matrices $\left\{ H_{12}\left(t\right),H_{11}\left(t\right)\right\} _{t=1}^{2}$
from phase 1, and also consequently of the transformation matrices
$\left\{ U_{1}(t)\right\} _{t=1}^{2}$. This allows $T_{2}$ to calculate
all the interference caused at $R_{1}$ in phase 1, namely $J_{1},J_{2}$
and $I_{1},I_{2},I_{3},I_{4}$. Now, if we can provide these $6$
symbols to $R_{2}$, it will be possible for $R_{2}$ to decode all
its intended DSs. Also, as mentioned earlier, knowledge of $I_{1},...,I_{4}$
is useful at $R_{1}$ allowing it to cancel interference from phase
1. The interference symbols $J_{1}$ and $J_{2}$ are already known
at $R_{1}$, and it can simply cancel them out from its received signal.
In other words, the interfering symbols $J_{1}$ and $J_{2}$ are
retrospectively aligned with the interference from phase 1 at $R_{1}$.

At time slot $t=3,$ $T_{2}$ transmits $J_{1}$ and $I_{1}$, while
$T_{1}$ transmits $u_{5},u_{6}$ again in the null-space of $H_{21}$.
Since $R_{2}$ sees no interference from $T_{1}$, it is able to determine
both $J_{1}$ and $I_{1}$. Receiver $R_{1}$ cancels outs the already
known $J_{1}$, and after using a unitary transformation, is able
to determine $I_{1}$ as well as its desired DSs $u_{5}$ and $u_{6}$.
It then cancels out $I_{1}$ from the combination $LC_{1}+I_{1}$
received at time $t=1$, to obtain the linear combination $LC_{1}$.
The same strategy is again used at time $t=4$, where $T_{2}$ transmits
$J_{2}$ and $I_{3}$ and $T_{1}$ transmits $u_{7},\, u_{8}$. $R_{1}$
obtains $u_{7}$, $u_{8}$ and $I_{3}$, and uses $I_{3}$ to cancel
interference from $LC_{3}+I_{3}$ it obtained at $t=2$ and thus learn
$LC_{3}$.

At $t=5$, $T_{1}$ needs to transmit only $u_{9}$, which it again
transmits in the null space of $H_{21}$. $T_{2}$ transmits the remaining
interference symbols $I_{2}$ and $I_{4}$. $R_{1}$ uses a unitary
transformation $U_{1}(t)$ such that only its first antenna is affected
by $u_{9}$. It uses the remaining two antennas to decode the two
symbols $I_{2}$ and $I_{4}$, which it then cancels from its received
signals obtained in Phase $1$ to obtain $LC_{2}$, $LC_{4}$ and
$u_{9}$. Thus, at the end of phase 2, $R_{1}$ has $4$ independent
linear combinations $LC_{1},...,LC_{4}$ of $u_{1},\dots,u_{4}$ and
the DSs $u_{5},\dots,u_{9}$, and thus can decode all its intended
data symbols. Similarly, $R_{2}$ has $5$ linear combinations of
$v_{1},\dots,v_{5}$, namely $LC^{1}$, $LC^{2}$, $I_{1}$, $I_{2}$
and $J_{1}$ and five linear combinations of $v_{6},\dots,v_{10}$,
namely $LC^{3}$, $LC^{4}$, $I_{3}$, $I_{4}$ and $J_{2}$. Hence
$R_{2}$ can decode all its desired data symbols $v_{1},\dots,v_{10}$.
The complete transmission scheme is clearly illustrated in Fig. \ref{fig:Alignment-Scheme-for}.
\begin{rem}
\label{More-constrained-achievability-scheme-remark}The careful reader
will have noticed that although $T_{1}$ has (instantaneous) knowledge
of the direct channel to its unpaired receiver i.e., $H_{22}$, it
never actually uses this information. The same holds true for the
generalized achievability scheme in Section \ref{sec:ACHIEVABILITY-SCHEME},
which makes the achievability scheme developed here also applicable
to the weaker version of the hybrid CSIT models.
\end{rem}

\begin{rem}
The final two columns of Table \ref{tab:Active-OUTER-BOUNDS} give
a concise comparison of the DoF regions for delayed CSIT, hybrid CSIT
1 and 2, and perfect and instantaneous CSIT. We notice that, except
for the case $M_{1},M_{2}>N_{1},N_{2}$, hybrid CSIT 1 achieves the
DoF region for perfect and instantaneous CSIT and the DoF region for
hybrid CSIT 2 is the same as that of delayed CSIT. It is only for
the Case A.I.3b in Table \ref{tab:Active-OUTER-BOUNDS}, i.e., $M_{1}>N_{1},\: M_{2}>N_{2},\: M_{2}>N_{1}$
that we observe the DoF region for both the hybrid CSIT models lying
between the DoF region with only delayed CSIT and the DoF region with
perfect and instantaneous CSIT. For the $\left(4,5,3,2\right)$ MIMO
IC example used earlier in this Section, after fixing $d_{2}=2$,
we can achieve the DoF tuple $\left(2,2\right)$ with perfect and
instantaneous CSIT, but only $\left(0,2\right)$ DoF with delayed
CSIT (although the corner point $\left(\frac{18}{7},\frac{5}{7}\right)$
is also achievable with delayed CSIT), and we recall that we were
able to achieve $\left(\frac{9}{5},2\right)$ DoF with hybrid CSIT
1. The difference between the DoF regions of the two hybrid CSIT models
can be explained by the fact that transmit beamforming in hybrid CSIT
1 allows the interference to be zero-forced at $R_{2}$, which being
the receiver with the fewer antennas is more constrained than $R_{1}$.
This allows hybrid CSIT 1 to use its instantaneous CSIT for a greater
gain.
\end{rem}

\subsection{An Alternating CSIT Example}

The alternating CSIT model allows for the CSIT configuration to change
between time slots e.g., between hybrid CSIT 1 and hybrid CSIT 2.
All the achievability schemes presented in \cite{DBLP:journals/corr/abs-1208-5071}
for the 2-user MISO BC can be adapted to the 2-user MISO IC, where
each transmitter has 2 antennas and each receiver has a single antenna.
The $3^{2}$ possible CSIT states in the MISO BC are mapped into the
$3^{8}$ CSIT states of the MISO IC as follows: CSIT at $T_{2}$ about
the cross-channel $H_{12}$ to $R_{1}$ in the IC is the same as the
CSIT about the channel to receiver 1 in the BC, and similarly, CSIT
about the channel to receiver 2 in the BC corresponds to CSIT at $T_{1}$
about the cross-channel $H_{21}$ to $R_{2}$, with neither transmitter
having any knowledge about the rest of the channels.  Here, we  further
develop the alternating CSIT model for the 2-user MIMO IC in this
example, elaborating a scheme that improves on the $\left(4,5,3,2\right)$-MIMO
IC hybrid CSIT 1 example given previously.

We use the freedom of alternating between the two hybrid CSIT models
and the synergistic gains thus obtained to demonstrate the achievability
of the DoF pair $\left(\frac{15}{8},2\right)$ for the same $\left(M_{1},M_{2},N_{1},N_{2}\right)=\left(4,5,3,2\right)$
IC of Section \ref{sec:ACHIEVABILITY-SCHEME}. We note that the $\left(\frac{15}{8},2\right)$
DoF pair achieved with alternating CSIT is strictly better than the
DoF tuple $\left(\frac{9}{5},2\right)$ achievable with just hybrid
CSIT 1 and the DoF tuples $\left(\frac{9}{5},\frac{11}{10}\right)$
and $\left(0,2\right)$ achievable with only hybrid CSIT 2. To achieve
this DoF pair, it is sufficient for $T_{1}$ to send $30$ DSs to
$R_{1}$ and $T_{2}$ to send $32$ DSs to $R_{2}$ over $16$ time
extensions. We do this by using the hybrid CSIT 1 model for the first
$15$ time slots, and shifting to hybrid CSIT 2 for the last time
slot. The DSs to be transmitted from $T_{1}$ are denoted as $u_{1},u_{2},\dots,u_{30}$
and the DSs transmitted from $T_{2}$ are denoted as $v_{1},v_{2},\dots,v_{32}$.
\begin{figure*}[t]
\includegraphics[scale=0.5]{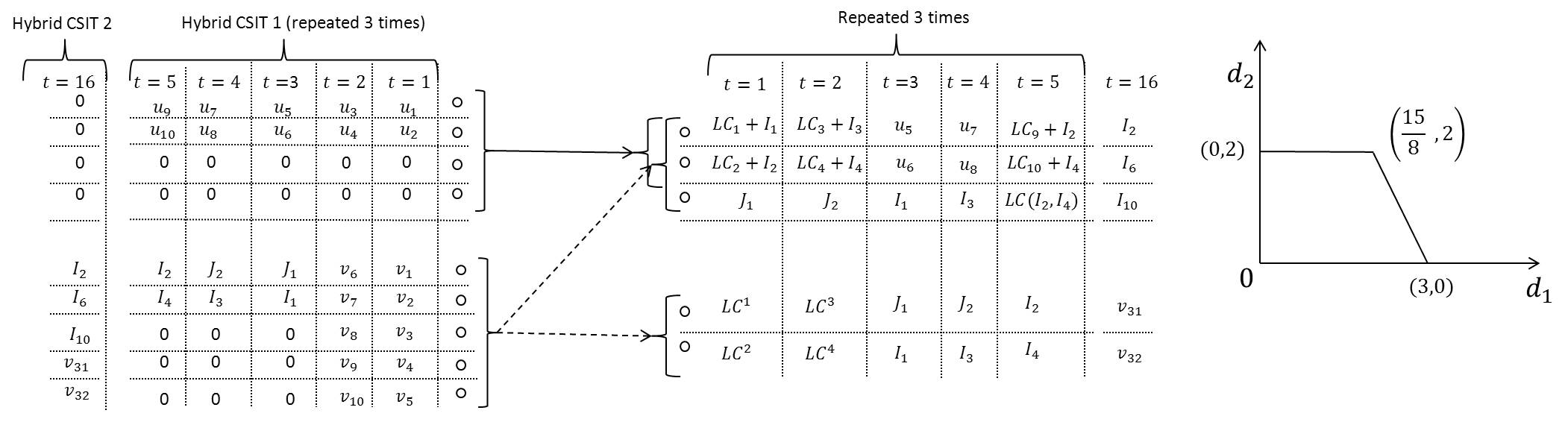}\caption{Interference alignment scheme for achieving $\left(\frac{15}{8},2\right)$
DoF pair over the $(M_{1},M_{2},N_{1},N_{2})=(4,5,3,2)$ MIMO IC with
alternating CSIT.\label{fig:Alignment-Scheme-Alternating-CSIT}}
\end{figure*}

\textbf{\uline{Hybrid CSIT 1 Phase}}

We ignore the additive noise henceforth, since it does not affect
the DoF region. We divide the total of 15 times slots with hybrid
CSIT 1 into $3$ similar phases of $5$ time slots each, and depict
one of these phases in Fig. \ref{fig:Alignment-Scheme-Alternating-CSIT}.
Each of these phases is similar to the hybrid CSIT 1 example explained
previously in Fig. \ref{fig:Alignment-Scheme-for}, including an unitary
matrix transformation $U_{1}(t)$ at $R_{1}$ to restrict the effect
of its desired DSs to only the first $2$ antennas. The difference
lies in the $5^{th}$ time slot, where $T_{1}$ now sends $2$ DSs,
$u_{9}$ and $u_{10}$ instead of a single DS. The transmission scheme
for $T_{2}$ remains the same for the $5$ time slots. At the end
of this phase, $R_{2}$ has 5 linear combinations of $v_{1},\dots,v_{5}$
i.e., $LC^{1},LC^{2},I_{1},I_{2}$ and $J_{1}$ and 5 linear combinations
of $v_{6},\dots,v_{10}$, namely $LC^{3},LC^{4},I_{3},I_{4}$ and
$J_{2}$, and is thus able to decode all its DSs $v_{1},\dots,v_{10}$.
On the other hand, $R_{1}$ is unable to learn the interference $I_{2}$
and $I_{4}$, and as a result it can not cancel the interference to
decode all of its desired DSs $u_{1},...,u_{10}$. We also observe
that $R_{1}$ knows one linear combination of $I_{2}$ and $I_{4}$
(at $t=5$), denoted as $LC(I_{2},I_{4})$ in Fig. \ref{fig:Alignment-Scheme-Alternating-CSIT},
and thus can learn $I_{4}$ if it is provided with $I_{2}$. This
shall allow it to cancel all the interference from its received outputs
at $t=1$, $t=2$ and $t=5$ to obtain four linear combinations, namely
$LC_{1},\dots,LC_{4}$ of $u_{1},\dots,u_{4}$, two linear combinations
$LC_{9},LC_{10}$ of $u_{9},u_{10}$ and the DSs $u_{5},\dots,u_{8}$,which
is sufficient information to decode all its desired DSs $u_{1},\dots,u_{10}$
from this phase. For this reason, providing $I_{2}$ to $R_{1}$ shall
be the goal of our hybrid CSIT 2 phase.

The above hybrid CSIT 1 phase is repeated $3$ times, for a total
of $15$ time slots. In these $15$ time slots, $T_{1}$ transmits
DSs $u_{1},\dots,u_{30}$ while $T_{2}$ transmits DSs $v_{1},\dots,v_{30}$.
$R_{2}$, as explained in the previous paragraph, is able to decode
its intended DSs $v_{1},\dots,v_{30}$. $R_{1}$ on the other hand,
as seen in the first phase, is unable to learn the interference symbols
$I_{2},I_{4}$ (from phase 1), $I_{6},I_{8}$(from phase 2) and $I_{10},I_{12}$
(from phase 3), and hence is unable to cancel out all the interference.
But since it possesses linear combinations $LC(I_{2},I_{4})$, $LC(I_{6},I_{8})$
and $LC(I_{10},I_{12})$, at the end of each of the three phases respectively,
knowledge of $I_{2},I_{6}$ and $I_{10}$ will allow it to determine
the remaining interference symbols $I_{4}$,$I_{8}$ and $I_{12}$.
This, in turn, allows $R_{1}$ to strip off all the interference from
its received outputs, as illustrated in the previous paragraph, to
learn sufficient number of linear combinations to be able to decode
all its desired DSs $u_{1},\dots,u_{30}$. Thus, the goal of the hybrid
CSIT 2 phase shall be to ensure that $R_{1}$ learns $I_{2}$, $I_{6}$
and $I_{10}$. In the same phase, $T_{2}$ also transmits two extra
DSs $v_{31}$ and $v_{32}$ intended for receiver $R_{2}$.

\textbf{\uline{Hybrid CSIT 2 Phase}}

At time slot $t=16$, we switch to the hybrid CSIT 2 model, where
$T_{2}$ has perfect and instantaneous CSIT about $H_{12}$. Since
$H_{12}$ is a $3\times5$ matrix, it has a $2$ dimensional null
space. $T_{2}$ transmits $2$ DSs $v_{31}$ and $v_{32}$ in this
null space, which are consequently not visible at $R_{1}$. Along
with $v_{31}$ and $v_{32}$, $T_{2}$ also re-transmits $I_{2},I_{6}$
and $I_{10}$, having learnt these interference symbols in the hybrid
CSIT 1 phases through delayed CSIT. $T_{1}$ does not transmit during
this phase. Since $R_{2}$ has already knows $I_{2},I_{6}$ and $I_{10}$
from the previous hybrid CSIT 1 phases, it is able to decode the two
intended DSs $v_{31}$ and $v_{32}$, which gives a total of $32$
DSs in $16$ time slots for $R_{2}$. Receiver $R_{1}$, which does
not see $v_{31}$ and $v_{32}$ because of the aforementioned transmit
beamforming at $T_{2}$, is able to learn the $3$ interference symbols
$I_{2},I_{6}$ and $I_{10}$ using its $3$ antennas, and uses these
with its earlier knowledge of $LC(I_{2},I_{4})$, $LC(I_{6},I_{8})$
and $LC(I_{10},I_{12})$ to learn the remaining interference symbols
$I_{4},I_{8}$ and $I_{12}$. Now $R_{1}$ can cancel away all the
interference symbols it has seen in the previous time slots, and thus
obtains sufficient linear combinations of its desired DSs to decode
$u_{1},...,u_{30}$ at the end of $16$ time slots. Thus, at the conclusion
of the hybrid CSIT 2 phase, we achieve the promised $\left(\frac{30}{16},\frac{32}{16}\right)=\left(\frac{15}{8},2\right)$
DoF pair, using alternating CSIT.

\section{Achievability Scheme\label{sec:ACHIEVABILITY-SCHEME}}

In this section, we describe an achievability scheme for the $\left(M_{1},M_{2},N_{1},N_{2}\right)$
MIMO IC, which is applicable for the case $M_{1},M_{2}>N_{1},N_{2}$,
for both hybrid CSIT models. The scheme is explained here in detail
for hybrid CSIT 1. Since the same scheme is easily adapted for the
hybrid CSIT 2 model by simply interchanging the roles played by the
two transmitters, the details for hybrid CSIT 2 will not be mentioned
explicitly.

The achievability scheme is a hybrid of transmit beamforming and retrospective
interference alignment, and is denoted henceforth as $\mathcal{HIA}$
(Hybrid Interference Alignment). It is a generalization of the example
described in the previous section. An overview of the scheme is given
below, with the full exposition following later in this section. 
\begin{enumerate}
\item $\mathcal{HIA}$ is designed to transmit $\left(d_{1}^{*},d_{2}^{*}\right)$
data symbols (DSs) by coding over $T$ time slots, thus achieving
$\left(d_{1},d_{2}\right)=\left(\frac{d_{1}^{*}}{T},\frac{d_{2}^{*}}{T}\right)$
DoF pair. The scheme is divided into two phases, consisting of $t_{1}$
and $t_{2}$ time slots respectively. 
\item The first phase is designed such that $T_{2}$ is able to transmit
all the $d_{2}^{*}$ DSs. At each time slot $t$ in the first phase,
$T_{2}$ transmits DSs using all its $M_{2}$ transmit antennas, while
$T_{1}$ transmits as many DSs as possible in the null space of the
channel $H_{21}$. Thus, transmit beamforming allows $T_{1}$ to null
out all interference at the receiver $R_{2}$.
\item Receive beamforming allows receiver $R_{1}$ to separate out the interference
from $T_{2}$ into two subsets, $S_{12}$ and $S_{2}$. These two
subsets correspond respectively to the subsets $I$ and $J$ from
the example in Section \ref{sub:Example-of-Achievability}. $R_{1}$
is able to decode the interference symbols from $S_{2}$, and hence
$S_{2}$ can be used by $T_{2}$ for retrospective interference alignment
in the next phase. The aim of the next phase is to provide $S_{12}$
at both receivers, and $S_{2}$ at receiver $R_{2}$. We will see
that this gives each receiver sufficiently many independent linear
combinations to be able to decode all of its intended DSs.
\item At each time slot $t$ in phase 2, $T_{1}$ transmits as much as possible
in the null space of $H_{21}$, while $T_{2}$ transmits a combination
of symbols from $S_{2}$ and $S_{12}$. As mentioned earlier, $S_{2}$
is retrospectively aligned with the interference already known at
$R_{1}$, and is easily canceled out. Thus, $R_{1}$ decodes the interference
symbols $S_{12}$, as a result of which the effects of interference
are canceled out from its received signals in phase 1. This allows
$R_{1}$ to isolate linear combinations of its intended DSs. $R_{2}$
is able to decode $S_{2}$ and $S_{12}$, which are linear combinations
of its intended DSs. As seen in the previous example, both receivers
are finally able to decode all their intended DSs from these linear
combinations.
\end{enumerate}
The details of the alignment scheme are explained below.\\
\uline{Phase 1 }We choose the number of time slots $t_{1}$ in
this phase such that all the needed $d_{2}^{*}$ DSs can be sent from
transmitter $T_{2}$, using all $M_{2}$ antennas. In the same phase,
at each time slot $t$, $T_{1}$ transmits $x=\min(M_{1}-N_{2},M_{2}-N_{2})$
DSs, taking care to keep them all in the null space of $H_{21}(t)$.
Hence, these $x$ DSs create no interference at $R_{2}$. 

$R_{1}$ then calculates a unitary transformation matrix $B(t)$ from
its knowledge of $H_{11}(t)$ such that the $x$ symbols from $T_{1}$
affect only the first $x$ antennas of $R_{1}$. Such a unitary transformation
always exists, being essentially a change of basis transformation
into a new orthonormal basis such that the first $x$ basis vectors
now span the $x-$dimensional received signal space at $R_{1}$, and
can be calculated using a singular value decomposition (see 7.4.3
in \cite{Horn1985}). The antennas denote the coordinates of the received
signal in this new basis, and hence only the first $x$ antennas have
non-zero symbols after the unitary transformation $B(t)$. The interference
at any antenna $i$ of $R_{1}$ in time slot $t$ is denoted as $I_{i}\left(t\right)$.
Interference at multiple antennas, e.g., at all antennas from $i$
to $j$ at time $t$ is denoted as $I_{i:j}\left(t\right)$. We next
divide the interference at $R_{1}$ into two parts. 

$S_{12}:=\{I_{1:x}(t)\}_{t=1}^{t_{1}}$, is the collection of all
interfering symbols at $R_{1}$ that have to be learnt by $R_{1}$
in phase 2 to be able to decode the $xt_{1}$ data symbols of phase
$1$. These interfering symbols will be known at $T_{2}$ in phase
2 because of delayed CSIT and will also help $R_{2}$ to decode its
intended data symbols by providing extra linear combinations of the
transmitted symbols. It is easily shown that all these interfering
symbols are linearly independent almost surely.

$S_{2}:=\{I_{[x+1:M_{2}-N_{2}]}(t)\}_{t=1}^{t_{1}}$is the interference
at $R_{1}$ that must be learned only at $R_{2}$ in phase 2. Since
these interfering symbols are already known at $R_{1}$ by the end
of phase 1, retrospective interference alignment is possible in phase
2. $T_{2}$ knows these interfering symbols in phase 2 because of
delayed CSIT. 

\uline{Phase 2 }The goals to be achieved at each transmitter/receiver
in phase 2 are as follows:
\begin{description}
\item [{$T_{1}$}] $T_{1}$ sends $d_{1}^{*}-xt_{1}$ new data symbols
over phase 2. As much as possible, these should be sent in the null
space of $H_{21}$ to minimize interference at $R_{2}$. The $\Delta(t)$
symbols, at each time slot $t$, that can not be sent in the null
space will cause interference at $R_{2}$. This gives rise to the
constraint 
\begin{equation}
d_{1}^{*}-xt_{1}\leq(M_{1}-N_{2})t_{2}+\sum_{t=t_{1}+1}^{t_{2}}\Delta(t).\label{eq:constraint 1}
\end{equation}

\item [{$R_{1}$}] $R_{1}$ should be able to accommodate both the $d_{1}^{*}-xt_{1}$
data symbols sent by $T_{1}$ as well as the interfering symbols $S_{12}$
sent by $T_{2}$. The symbols from $S_{2}$ are already known and
can be discarded. This gives rise to the constraint
\begin{eqnarray}
N_{1}t_{2}-(d_{1}^{*}-xt_{1}) & \geq & |S_{12}|=xt_{1}\nonumber \\
\text{i.e.,\,\,\ }d_{1}^{*} & \leq & N_{1}t_{2}.\label{eq:constraint 2}
\end{eqnarray}

\item [{$R_{2}$}] $R_{2}$ faces $\sum\Delta(t)$ interference symbols
from $T_{1}$ which it discards. $R_{2}$ must also be able to decode
both $S_{12}$ and $S_{2}$. This gives rise to the constraint
\begin{eqnarray}
 &  & |S_{12}|+|S_{2}|+\sum_{t=t_{1}+1}^{t_{2}}\Delta(t)\leq N_{2}t_{2}\nonumber \\
 &  & (M_{2}-N_{2}-x)t_{1}+xt_{1}+\sum_{t=t_{1}+1}^{t_{2}}\Delta(t)\leq N_{2}t_{2}\nonumber \\
 &  & \implies(M_{2}-N_{2})t_{1}\leq N_{2}t_{2}-\sum_{t=t_{1}+1}^{t_{2}}\Delta(t).\label{eq:constraint 3}
\end{eqnarray}

\item [{$T_{2}$}] $T_{2}$ must send both $S_{12}$ (which is needed at
both $R_{1}$ and $R_{2}$) and $S_{2}$ (which aligns with interference
already known at $R_{1}$). At any given time slot $t$, $T_{2}$
transmits according to the following constraints, \textbf{(i)} no
more than $N_{2}-\Delta(t)$ symbols are transmitted, which ensures
that $R_{2}$ can decode the symbols after discarding the $\Delta(t)$
interference symbols and \textbf{(ii)} no more than $N_{1}$ elements
of $S_{2}\bigcup S_{12}$ are sent, allowing $R_{1}$ to accommodate
these $N_{1}$ symbols.
\end{description}
Thus, this achievability scheme can to send $(d_{1}^{*},d_{2}^{*})$
symbols for the $(M_{1},M_{2},N_{1},N_{2})$ over $T=t_{1}+t_{2}$
time slots, provided the constraints \eqref{eq:constraint 1}- \eqref{eq:constraint 3}
are satisfied. We shall prove later in Section \ref{sub:Proof-of-Theorem-hybrid CSIT 1}
that only Case A.I.3b from Table \ref{tab:Active-OUTER-BOUNDS} requires
our attention, and show that the achievability schemes for delayed
CSIT and perfect and instantaneous CSIT suffice for the rest of the
cases. We divide Case A.I.3b again into 3 mutually exclusive and exhaustive
sub-cases, which are shown to have different DoF regions in Section
\ref{sec:Proofs-of-Main-Results}, which are defined below,
\begin{description}
\item [{Case}] \textbf{I: }$M_{2}\leq M_{1}$,
\item [{Case}] \textbf{II: }$M_{1}<M_{2}$,\ $N_{1}\left(M_{2}-N_{2}\right)\leq M_{2}\left(M_{1}-N_{2}\right)$,
\item [{Case}] \textbf{III: }$M_{1}<M_{2}$,\ $N_{1}\left(M_{2}-N_{2}\right)>M_{2}\left(M_{1}-N_{2}\right).$
\end{description}
We next check if these constraints are satisfied for the above sub-cases\textbf{. }

\textbf{CASE I} The parameters for Case I are $d_{1}^{*}=N_{1}(M_{2}-N_{2}),\ d_{2}^{*}=M_{2}N_{2},\ T=M_{2},$
$t_{1}=N_{2},\ t_{2}=M_{2}-N_{2},\ x=M_{2}-N_{2},\ \Delta(t)=0$,
and inequality \eqref{eq:constraint 1} is now equivalent to $N_{1}(M_{2}-N_{2})-(M_{2}-N_{2})N_{2}\leq(M_{1}-N_{2})(M_{2}-N_{2}),$
which is true since $N_{1}-N_{2}\leq M_{1}-N_{2}$ is true under Case
I. It is also easily verified that inequalities \eqref{eq:constraint 2}
and \eqref{eq:constraint 3} hold (with equality) for the parameters
of Case I as well. 

\textbf{CASE II} The parameters for Case II are $d_{1}^{*}=N_{1}(M_{2}-N_{2}),\ d_{2}^{*}=M_{2}N_{2},\ T=M_{2},\ t_{1}=N_{2},\ t_{2}=M_{2}-N_{2},\ x=M_{1}-N_{2},\ \Delta(t)=0$,
for which inequality \eqref{eq:constraint 1} is equivalent to $N_{1}(M_{2}-N_{2})-(M_{1}-N_{2})N_{2}\leq(M_{1}-N_{2})(M_{2}-N_{2}),$
which is true since $N_{1}(M_{2}-N_{2})\leq M_{2}(M_{1}-N_{2})$ is
true under Case II. Similarly, it is easily verified that inequalities
\eqref{eq:constraint 2} and \eqref{eq:constraint 3} hold (with equality)
for the parameters of Case I as well.

\textbf{CASE III} The DoF region for Case III has two non-trivial
corner points, of which the point $\left(\frac{N_{1}\left(M_{2}-M_{1}\right)}{M_{2}-N_{1}},\frac{M_{2}\left(M_{1}-N_{1}\right)}{M_{2}-N_{1}}\right)$
is considered first. For this point, the parameters are $d_{1}^{*}=N_{1}(M_{2}-M_{1})\ ,d_{2}^{*}=M_{2}\left(M_{1}-N_{1}\right),\ T=M_{2}-N_{1},\ t_{1}=M_{1}-N_{1},\ t_{2}=M_{2}-M_{1},\ x=M_{1}-N_{2}$.
Choose $\Delta(t)$ such that all the empty dimensions at $R_{2}$
in phase 2 are utilized i.e., $\sum_{t=t_{1}+1}^{t_{2}}\Delta(t)=N_{2}t_{2}-(M_{2}-N_{2})t_{1}$,
and we thus obtain $\sum_{t=t_{1}+1}^{t_{2}}\Delta(t)=N_{1}\left(M_{2}-N_{2}\right)-M_{2}\left(M_{1}-N_{2}\right).$With
these parameters, inequality \eqref{eq:constraint 1} is equivalent
to 
\[
N_{1}(M_{2}-M_{1})-(M_{1}-N_{2})(M_{1}-N_{1})\leq(M_{1}-N_{2})(M_{2}-M_{1})+N_{1}(M_{2}-N_{2})-M_{2}(M_{1}-N_{2}),
\]
 which holds with equality for the $1^{st}$ corner point Case III.
Inequalities \eqref{eq:constraint 2} and \eqref{eq:constraint 3}
are similarly verified to hold (with equality) for this corner point.We
now consider the $2^{nd}$ non-trivial corner point of the DoF region
under Case III i.e, $\left(M_{1}-N_{2},N_{2}\right)$, for which the
parameters are $d_{1}^{*}=(M_{1}-N_{2})M_{2},\ d_{2}^{*}=N_{2}M_{2},\ T=M_{2},\ t_{1}=N_{2},\ t_{2}=M_{2}-N_{2},\ x=M_{1}-N_{2},\ \Delta(t)=0$.
Inequality \eqref{eq:constraint 1} is now equivalent to $(M_{1}-N_{2})M_{2}-(M_{1}-N_{2})N_{2}\leq(M_{1}-N_{2})(M_{2}-N_{2}),$
which is true. Similarly, inequalities \eqref{eq:constraint 2} and
\eqref{eq:constraint 3} are easily verified to hold (with equality)
for this corner point.

We find that all the constraints are satisfied, and thus the achievability
scheme can accommodate all the possible sub-cases.

\section{Proofs of Main Results \label{sec:Proofs-of-Main-Results}}

\subsection{Proof of Theorem \ref{Theorem 5}\label{sub:Proof-of-Theorem-hybrid CSIT 1}}

Consider each of the sub-cases in Table \ref{tab:Active-OUTER-BOUNDS}
for the hybrid CSIT 1 model.
\begin{description}
\item [{0}] As seen in Table \ref{tab:Active-OUTER-BOUNDS}, the only active
outer bound for the delayed CSIT case is the perfect and instantaneous
bound $L_{3}$. This outer bound is achievable even with just delayed
CSIT, and therefore we have $D^{d}=D^{h1}=D^{i}$.
\item [{B}] Since $M_{2}<N_{1}$, there is no null space from transmitter
$T_{2}$ to $R_{1}$. This allows us to use the transmit beamforming
achievability scheme from \cite{DBLP:journals/tit/JafarF07}, for
the case where both transmitters have perfect and instantaneous CSIT.
$T_{2}$ is supposed to transmit in the null space of $H_{12}$; but
since no such null space exists for this case, it does not matter
for the scheme whether $T_{2}$ has knowledge of channel state or
not. This allows us to achieve the complete DoF region for the perfect
and instantaneous case with hybrid CSIT 1. Thus, $D^{h1}=D^{i}$.
\item [{A.II}] Since $M_{2}<N_{1}$, we use the same argument as for case
B to prove that $D^{h1}=D^{i}$.
\item [{A.I.1}] The only outer bound that holds in this case is $L_{3}\equiv d_{1}+d_{2}\leq M_{1}.$The
outer bound region is given in Fig. \ref{fig:1AI1}. It is clear that
it can be easily achieved by simple zero-forcing at transmitter $T_{1}$.
Thus, $D^{h1}=D^{i}$.
\item [{A.I.2}] As calculated previously in the proof of Theorem \ref{Theorem 4},
the outer bounds $L_{2}$ and $L_{3}$ coincide. This shows that the
DoF region for this case is the same as $D^{i}$, which is achievable
even by the interference alignment scheme for delayed CSIT. Thus,
$D^{h1}=D^{i}$.
\item [{A.I.3a}] We consider first the sub-case of A.I.3 where $M_{2}=N_{1}$.
Calculating the outer bounds, we see that $L_{2}$ and $L_{3}$ coincide
and are given as $d_{1}+d_{2}\leq M_{2}.$ The DoF outer bound region
is given by Fig. \ref{fig:1A13a}. It is clear that this DoF region
can be easily achieved by transmit beamforming in the null space of
$H_{21}$ at transmitter $T_{1}$. Hence, $D^{h1}=D^{i}$.
\begin{figure}[tb]
\raggedright{}%
\begin{minipage}[t]{0.45\textwidth}%
\begin{center}
\includegraphics[clip,scale=0.45]{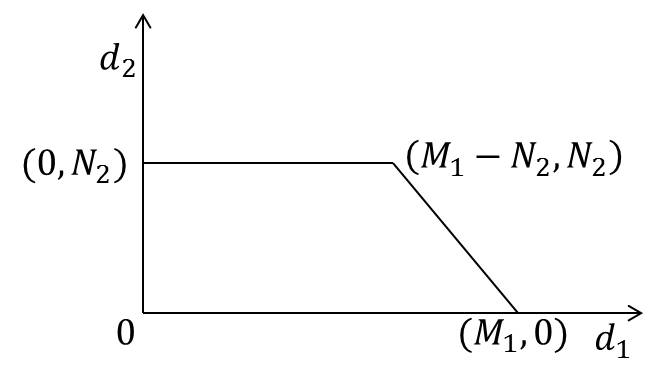}
\par\end{center}

\begin{center}
\caption{\label{fig:1AI1}Case A.I.1 (Hybrid CSIT 1)}

\par\end{center}%
\end{minipage}\hfill{}%
\begin{minipage}[t]{0.45\textwidth}%
\begin{center}
\includegraphics[clip,scale=0.45]{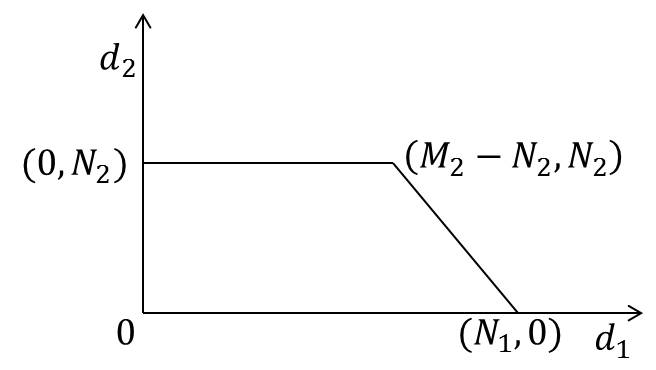}
\par\end{center}

\begin{center}
\caption{\label{fig:1A13a}Case A.I.3.a (Hybrid CSIT 1)}

\par\end{center}%
\end{minipage}
\end{figure}

\item [{A.I.3b}] We calculate the applicable outer bounds for this case
namely when $M_{1},M_{2}>N_{1},N_{2}$, which are 
\begin{eqnarray*}
L_{2} & \equiv & \frac{d_{1}}{N_{1}}+\frac{d_{2}}{\min(N_{1}+N_{2},M_{2})}\leq1;\,\,\, L_{3}\equiv d_{1}+d_{2}\leq\min(N_{1}+N_{2},M_{1},M_{2}).
\end{eqnarray*}
It is obvious that this is the DoF region that needs considerable
attention. It is convenient to further sub-divide this case into various
mutually exclusive and exhaustive sub-cases and consider each sub-case
separately.\\
When $N_{1}+N_{2}<M_{2}$, we can transmit at most with $\min(N_{1}+N_{2},M_{2})$
antennas without loss of DoF optimality and hence can always remove
any extra antennas from $T_{2}$ without affecting the DoF region.
Thus, we only need to address the sub-case where $N_{1}+N_{2}\geq M_{2}$.
The outer bounds now become 
\begin{eqnarray*}
L_{2} & \equiv & \frac{d_{1}}{N_{1}}+\frac{d_{2}}{M_{2}}\leq1;\,\,\, L_{3}\equiv d_{1}+d_{2}\leq\min(M_{1},M_{2}).
\end{eqnarray*}
This again gives rise to various sub-cases, which are enumerated below.\\
\textbf{Case I:} Recall that under Case I, $M_{2}\leq M_{1}$. The
outer bounds in this case are\textbf{ }
\begin{eqnarray*}
L_{2} & \equiv & \frac{d_{1}}{N_{1}}+\frac{d_{2}}{M_{2}}\leq1;\,\,\, L_{3}\equiv d_{1}+d_{2}\leq M_{2}.
\end{eqnarray*}
We find that the only active outer bound is $L_{2}$, and the DoF
outer bound region is shown in Fig. \ref{fig:1IandII}.\\
\textbf{Case II: }Recall that under Case II, \textbf{$M_{1}<M_{2}$,
$N_{1}(M_{2}-N_{2})\leq M_{2}(M_{1}-N_{2})$.} The outer bounds in
this case are 
\begin{eqnarray*}
L_{2} & \equiv & \frac{d_{1}}{N_{1}}+\frac{d_{2}}{M_{2}}\leq1;\,\,\, L_{3}\equiv d_{1}+d_{2}\leq M_{1}.
\end{eqnarray*}
Now, we find that $L_{3}$ is inactive if the following inequality
holds 
\begin{eqnarray}
 & \left(1-\frac{N_{2}}{M_{2}}\right)N_{1}\leq & M_{1}-N_{2}\nonumber \\
\text{i.e.,} & N_{1}(M_{2}-N_{2})\leq & M_{2}(M_{1}-N_{2})\label{eq: case AIIIB-2}
\end{eqnarray}
in which case, the DoF outer bound region is again as shown in Fig.
\ref{fig:1IandII}. \\
\textbf{Case III:} Recall that under Case III, $M_{1}<M_{2}$, $N_{1}(M_{2}-N_{2})>M_{2}(M_{1}-N_{2})$.
This is the sub-case where both the bounds $L_{2}$ and $L_{3}$ are
active. The condition for both the bounds to be active is
\begin{equation}
N_{1}(M_{1}-N_{2})\geq M_{2}(M_{1}-N_{2})
\end{equation}
and the DoF outer bound region is shown in Fig. \ref{fig:1III}.
\begin{figure}[tb]
\raggedright{}%
\begin{minipage}[t]{0.45\textwidth}%
\begin{center}
\includegraphics[clip,scale=0.45]{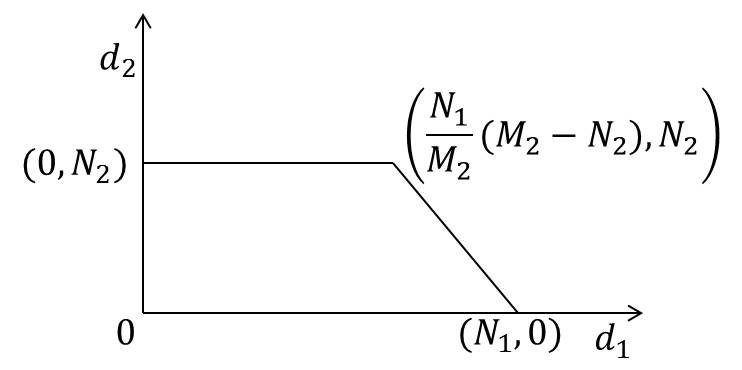}
\par\end{center}

\begin{center}
\caption{\label{fig:1IandII}Case I and II (Hybrid CSIT 1)}

\par\end{center}%
\end{minipage}\hfill{}%
\begin{minipage}[t]{0.45\textwidth}%
\begin{center}
\includegraphics[clip,scale=0.45]{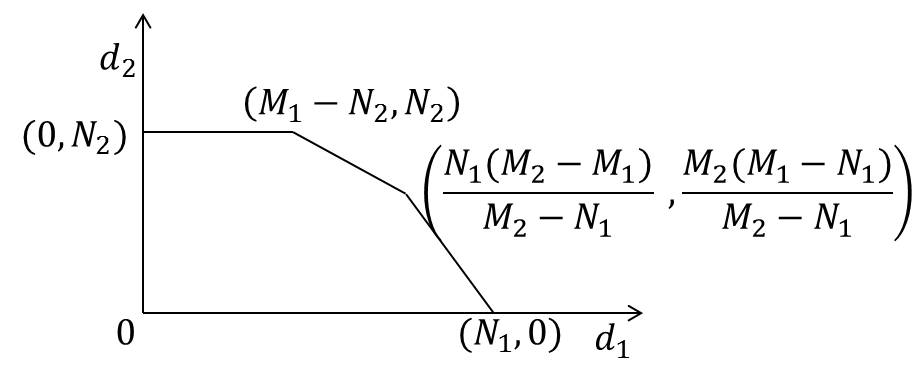}
\par\end{center}

\begin{center}
\caption{\label{fig:1III}Case III (Hybrid CSIT 1)}

\par\end{center}%
\end{minipage}
\end{figure}

\end{description}

\subsection{Proof of Theorem \ref{Theorem 4}}

Here, we analyze each of the sub-cases in Table \ref{tab:Active-OUTER-BOUNDS},
for the hybrid CSIT 2 model.
\begin{description}
\item [{0}] Since the only active outer bound is the perfect and instantaneous
bound $L_{3}$, which is achievable even with delayed CSIT, we have
$D^{h2}=D^{d}$.
\item [{B}] The DoF region for delayed CSIT is defined by outer bounds
$L_{\{1,3,4\}}$ , by Lemma \ref{Lemma 1}, which are also active
for the hybrid CSIT 2 case . Since this region is achievable even
under the more stringent delayed CSIT condition, we have the result
$D^{h2}=D^{d}$.
\item [{A.II}] In this case, the DoF region with delayed CSIT is defined
by the outer bounds $L_{\{1,3\}}$, which are also active for the
hybrid CSIT 2 case, by Lemma \ref{Lemma 1}. Since this region is
achievable even under the more stringent delayed CSIT condition, we
have the result $D^{h2}=D^{d}$.
\item [{A.I.1}] In case A.I.1, the DoF region with delayed CSIT is defined
by the outer bound $L_{1}$, by Lemma \ref{Lemma 1}, which is also
applicable for the hybrid CSIT 2 case. Since this region is already
achievable by the more stringent delayed CSIT condition, we have the
result $D^{h2}=D^{d}$.
\item [{A.I.2}] In this case, we have the conditions $M_{1}>N_{1},N_{2}$
and $M_{2}=N_{1}=N_{2}$. The outer bounds $L_{2}$ and $L_{3}$ now
coincide and become $d_{1}+d_{2}\leq M_{2}$. As a result, the only
active outer bound for delayed CSIT is $L_{3}$, by Lemma \ref{Lemma 3},
which is also active for the hybrid CSIT 2 case. Thus, by the same
argument as for the previous cases, we have the result $D^{h2}=D^{d}$.
\item [{A.I.3a}] We consider first the sub-case of A.I.3 where $M_{2}=N_{1}$.
By the same argument as for case A.I.2, we have $D^{h2}=D^{d}$.
\item [{A.I.3b}] This interesting case is defined by $M_{1},M_{2}>N_{1},N_{2}$.
The relevant outer bounds are 
\begin{eqnarray*}
L_{1} & \equiv & \frac{d_{1}}{\min(N_{1}+N_{2},M_{1})}+\frac{d_{2}}{N_{2}}\leq1;\,\,\, L_{3}\equiv d_{1}+d_{2}\leq\min(M_{1},M_{2}).
\end{eqnarray*}
When $N_{1}+N_{2}<M_{1}$, we can at most transmit with $\min(N_{1}+N_{2},M_{1})$
antennas and hence can always remove any extra antennas from $T_{1}$
without affecting the DoF region. Thus, we only need to discuss the
sub-case where $N_{1}+N_{2}\geq M_{1}$. The outer bounds now become
\begin{eqnarray*}
L_{1} & \equiv & \frac{d_{1}}{M_{1}}+\frac{d_{2}}{N_{2}}\leq1;\,\,\, L_{3}\equiv d_{1}+d_{2}\leq\min(M_{1},M_{2}).
\end{eqnarray*}
This again gives rise to cases I, II and III, defined previously in
Section \ref{sub:Proof-of-Theorem-hybrid CSIT 1}, which are analyzed
below.\\
\textbf{Case I: }Recall that under Case I, \textbf{$M_{1}\leq M_{2}$}.
The outer bounds in this case are\textbf{ }
\begin{eqnarray*}
L_{1} & \equiv & \frac{d_{1}}{M_{1}}+\frac{d_{2}}{N_{2}}\leq1;\,\,\, L_{3}\equiv d_{1}+d_{2}\leq M_{1}.
\end{eqnarray*}
We find that the only active outer bound is $L_{2}$, and the DoF
outer bound region is shown in Fig. \ref{fig:2IandII}. \\
\textbf{Case II:} Recall that under Case II, \textbf{$M_{2}<M_{1}$,
$N_{2}(M_{1}-N_{1})\leq M_{1}(M_{2}-N_{1})$}. The outer bounds in
this case are 
\begin{eqnarray*}
L_{1} & \equiv & \frac{d_{1}}{M_{1}}+\frac{d_{2}}{N_{2}}\leq1;\,\,\, L_{3}\equiv d_{1}+d_{2}\leq M_{2}.
\end{eqnarray*}
We find that $L_{3}$ is inactive under this case. Hence, the DoF
outer bound region is again as shown in Fig. \ref{fig:2IandII}.\\
\textbf{Case III:} Recall that under Case III, $M_{2}<M_{1}$, $N_{2}(M_{1}-N_{1})>M_{1}(M_{2}-N_{1})$.
This is the sub-case where both the bounds $L_{2}$ and $L_{3}$ are
active. Hence, the DoF outer bound region has the form given in Fig.
\ref{fig:2III}.
\begin{figure}[tb]
\begin{minipage}[t]{0.45\textwidth}%
\begin{center}
\includegraphics[clip,scale=0.45]{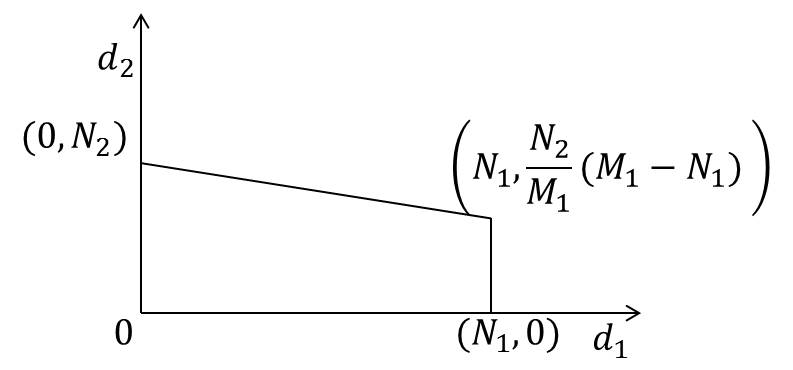}
\par\end{center}

\begin{center}
\caption{\label{fig:2IandII}Case I and II (Hybrid CSIT 2)}

\par\end{center}%
\end{minipage}\hfill{}%
\begin{minipage}[t]{0.45\textwidth}%
\begin{center}
\includegraphics[clip,scale=0.45]{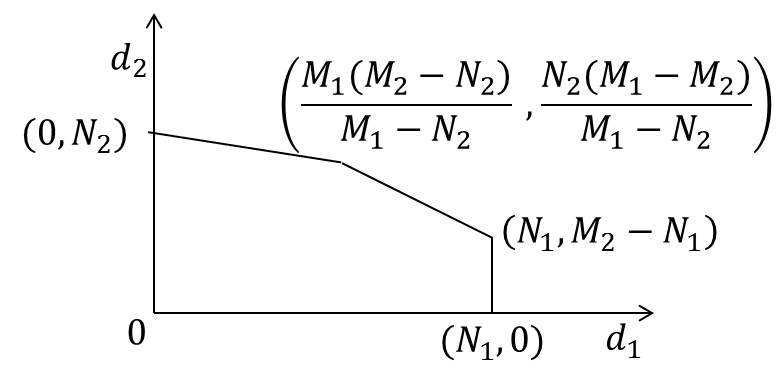}
\par\end{center}

\begin{center}
\caption{\label{fig:2III}Case III (Hybrid CSIT 2)}

\par\end{center}%
\end{minipage}
\end{figure}

\end{description}

\section{Conclusion\label{sec:CONCLUSION}}

In this paper, we study hybrid CSIT models for the general $(M_{1},M_{2},N_{1},N_{2})$
2-user interference channel, where it is assumed that one transmitter
has perfect and instantaneous CSIT of the channels at its unpaired
receiver while the other transmitter has delayed CSIT of the channels
at its unpaired receiver, with no channel state information at either
transmitter about the incoming channels at their respective paired
receivers. We obtain both DoF outer bounds and inner bounds, while
developing a new DoF-optimal interference alignment scheme tailored
for the hybrid CSIT models. The inner and outer bounds are shown to
coincide, and we are thus able to characterize the complete DoF regions
for both the hybrid CSIT models. By demonstrating an achievable scheme
that uses less information than in the nominal hybrid CSIT models
and by establishing that the DoF outer bounds hold for enhanced MIMO
ICs with more side information, we show that the DoF regions of the
nominal hybrid CSIT models actually apply to a total of $2\times2\times3^{5}$
hybrid CSIT models. We also demonstrate synergistic benefits from
switching between hybrid CSIT models by showcasing achievability schemes
that switch between the two hybrid CSIT models and achieve DoF tuples
beyond the DoF region of either hybrid CSIT model.

\bibliographystyle{IEEEtran}
\bibliography{bibliography}

\end{document}